\documentclass[twocolumn]{aastex701}
\usepackage{float}
\usepackage{xcolor}
\newcommand{\sigsfr}[0]{Log $\Sigma_{SFR}$ }
\newcommand{\sigmstar}[0]{Log $\Sigma_{M_*} $}
\usepackage{makecell}
\renewcommand\theadalign{bc}

\usepackage{hyperref}
\usepackage{makecell}
\renewcommand\theadalign{bc}
\usepackage{subfigure}
\usepackage{subcaption}

\begin{document}

\title{Searching Within Galaxies for the Earliest Signs of Quenching With Spatially Resolved Star Formation Histories in UVCANDELS Galaxies at z$<$ 0.3
}

 \correspondingauthor{Charlotte Olsen}

\author[0000-0002-8085-7578]{Charlotte Olsen}
\email{charlotte.olsen47@citytech.cuny.edu}
\affiliation{ 
Department of Physics, New York City College of Technology \textbf{},  
Brooklyn, NY 11201, USA}

\author[0000-0003-1530-8713]{Eric Gawiser}  
\email{gawiser@physics.rutgers.edu}
\affiliation{ 
Department of Physics and Astronomy, Rutgers University,  
Piscataway, NJ 08854, USA}

\author[0000-0001-5576-0144]{Charlotte Welker}
\email{CWelker@citytech.cuny.edu}
\affiliation{ 
Department of Physics, New York City College of Technology \textbf{},  
Brooklyn, NY 11201, USA}

\author[0000-0002-7064-5424]{Harry Teplitz}
\email{hit@ipac.caltech.edu}
\affiliation{California Institute of Technology: Pasadena, CA, US}

\author[0000-0001-9298-3523]{Kartheik Iyer}
\email{0000-0001-9298-3523}
\affiliation{Flatiron Institute Center for Computational Astrophysics}

\author[0000-0002-9373-3865]{Xin Wang}
\email{xwang@ucas.ac.cn}
\affiliation{School of Astronomy and Space Science, University of Chinese Academy of Sciences (UCAS), Beijing 100049, China}
\affiliation{National Astronomical Observatories, Chinese Academy of Sciences, Beijing 100101, China}
\affiliation{Institute for Frontiers in Astronomy and Astrophysics, Beijing Normal University, Beijing 102206, China}

\author[0000-0002-9946-4731]{Marc Rafelski}
\email{mrafelski@stsci.edu}
\affiliation{Space Telescope Science Institute, 3700 San Martin Drive, Baltimore, MD 21218, USA}

\author[0000-0001-8156-6281]{Rogier A. Windhorst} 
\email{Rogier.Windhorst@gmail.com}
\affiliation{School of Earth and Space Exploration, Arizona State University,
Tempe, AZ 85287-6004, USA}

\author[0000-0002-6610-2048]{Anton Koekemoer}
\email{koekemoer@stsci.edu}
\affiliation{Space Telescope Science Institute, 3700 San Martin Drive, Baltimore, MD 21218, USA}

\author[0000-0002-8630-6435]{Anahita Alavi}
\email{anahita@ipac.caltech.edu}
\affiliation{California Institute of Technology: Pasadena, CA, US}

\author[0000-0003-3759-8707]{Ben Sunnquist}
\email{bsunnquist@stsci.edu}
\affiliation{Space Telescope Science Institute, 3700 San Martin Drive, Baltimore, MD 21218, USA}

\author[0000-0001-9440-8872]{Norman Grogin}
\email{nagrogin@stsci.edu}
\affiliation{Space Telescope Science Institute, 3700 San Martin Drive, Baltimore, MD 21218, USA}

\author[0000-0003-2775-2002]{Yicheng Guo}
\email{guoyic@missouri.edu}
\affiliation{Department of Physics and Astronomy, University of Missouri, Columbia, MO, 65211, USA}

\author[0000-0002-2079-7438]{Christopher J. Conselice}
\email{0000-0002-2079-7438}
\affiliation{Jodrell Bank Centre for Astrophysics, University of Manchester, Oxford Road, Manchester UK}

\author[0000-0003-3466-035X]{{L. Y. Aaron} {Yung}}
\email{yung@stsci.edu}
\affiliation{Space Telescope Science Institute, 3700 San Martin Drive, Baltimore, MD 21218, USA}

\author[0000-0001-5294-8002]{Kalina Nedkova}
\email{knedkov1@jhu.edu}
\affiliation{Department of Physics and Astronomy, Johns Hopkins University, 3400 North Charles Street, Baltimore, MD 21218, USA}

\author[0000-0002-0108-4176]{Bahram Mobasher}
\email{mobasher@ucr.edu}
\affiliation{Department of Physics and Astronomy
University of California
Riverside, CA 92521, USA}

\author[0000-0003-1581-7825]{Ray A. Lucas}
\email{lucas@stsci.edu}
\affiliation{Space Telescope Science Institute, 3700 San Martin Drive, Baltimore, MD 21218, USA}

\author[0000-0001-7166-6035]{Vihang Mehta}
\email{vmehta@ipac.caltech.edu}
\affiliation{PAC, Mail Code 314-6, California Institute of Technology, 1200 E. California Boulevard, Pasadena, CA 91125, USA}

\author[0000-0002-7928-416X]{Y. Sophia Dai}
\email{0000-0002-7928-416X}
\affiliation{Chinese Academy of Sciences South America Center for Astronomy (CASSACA), \\
National Astronomical Observatories of China (NAOC),
CAS, 20A Datun Road, Beijing 100012, China}

\author[0000-0003-2098-9568]{Jonathan P. Gardner}
\email{jonathan.p.gardner@nasa.gov}
\affiliation{Sciences and Exploration Directorate, NASA Goddard Space Flight Center, 8800 Greenbelt Rd, Greenbelt, MD 20771, USA}

\begin{abstract}
Understanding the complicated processes that regulate star formation and cause a galaxy to become quiescent is key to our comprehension of galaxy evolution. We used eight well resolved star-forming z$<$ 0.3 galaxies from the UVCANDELS survey, where a total of 10 HST bands including UV follow up in UVIS/F275W allow us to reconstruct the star formation histories (SFHs) of regions across each galaxy. This approach provides a powerful tool to explore the spatio-temporal connection between star formation and galaxy evolution. The spatial and temporal profiles of stellar mass and star formation rate surface density were obtained from the SFHs of these regions. We measure scaling relations and projected radial profiles of regions within each galaxy at the time of observation and at 1 Gyr lookback time, noting possible trends in the evolution. By comparing the change in star formation over time we can infer the timing and location of star formation and see early signs of star formation shut off before quenching occurs. We compared the star formation rate density -- stellar mass density scaling relations for individual galaxies as they evolve from 1 Gyr lookback time. The correlation lines pivot around a log-stellar mass surface density of 7.25 [$M_\odot$ $kpc^{-2}$] may be evidence of a self-regulating process on these scales. Radial profiles of galaxy Log sSFR show an overall decrease over 1 Gyr, but five galaxies show a greater change in Log sSFR at the outskirts than the center indicating a possible early onset of quenching in these galaxies.

\end{abstract}



\section{Introduction}

The conversion of cold gas into stellar mass via star formation is a key process in galaxy formation. Understanding the rate at which a galaxy is currently forming stars, how much stellar mass it has built up, and determining how it has built up this stellar mass over time are all key to understanding galaxy evolution. 

However, the conversion of gas into stars is complicated by a number of other processes that regulate star formation and act on a range of scales. Globally, star formation can be fueled by cool gas channeled into the galaxy from the surrounding environment \citep{Lilly2013, Werk2016ApJ}, but galaxies in dense environments can have their star formation quenched when their gas is stripped or heated from ram pressure or mergers \citep{Gunn1972, Nulsen1982, Fujita1999,Bekki1999,Balogh2000,Bekki2002, Fujita2004}. Internal mechanisms for quenching can be seen on smaller scales, though, as feedback from stars, supernovae, and active galactic nuclei (AGN) can heat gas locally and suppress star formation \citep{Kimm2009, Hopkins2014, AnglsAlczar2017, ElBadry2017}. Star formation can also cease due to ``starvation'' \citep[e.g.][]{Larson1980, Balogh1997} where a galaxy is cut off from its gas supply, ``overconsumption'' where an abundance of star formation can drive outflows \citep{McGee2014} or ram pressure stripping \citep[see:][]{Guo2017, Guo2021}. Star formation has also been shown to be spatially dependent within the galaxy, with galaxies exhibiting age gradients that suggest either ``inside-out'' or ``outside-in'' growth \citep{Carrasco2010, vanDokkum2013, Pan2015, Zewdie2020} with additional evidence that star formation can be tied to the structure of the galaxy itself \citep[e.g.][]{Whittaker2015}. Quantifying and classifying the stochasticity of star formation caused by these smaller scale processes has helped to place tighter constraints on the timescales upon which they occur \citep{Broussard2019, FloresVelzquez2020, Broussard2022}. These processes are difficult to disentangle but occur on differing timescales \citep[e.g., ][]{Iyer2020}, which can in turn be observed via changes in the galaxy's star formation rate (SFR) over time, i.e. its star formation history (SFH)\citep{Carnall2019, Leja2019}.

Broadly, the combination of large- and small scale processes results in global properties and scaling relations that describe populations of galaxies at a given redshift. A common relation is that of SFR and stellar mass($M_*$). In $SFR-M_*$ space, star forming galaxies follow a roughly linear sequence, with quiescent galaxies lying below this relation. This correlation between SFR and $M_*$ in star forming galaxies is often called the star forming main sequence (SFMS)\citep[e.g.][]{Brinchmann2004, Daddi2007, Elbaz2007, Noeske2007, Salim2007, Peng2010, Speagle2014, Lee2018}. At a given redshift, the $SFR-M_*$ correlation  predicts the SFR of a star forming galaxy from its stellar mass, or vice versa. Starbursting galaxies lie above the correlation, with transitional or ``green valley'' galaxies lying between the star forming galaxies and the quiescent ``red and dead'' galaxies far below. Currently, new surveys  are working to fill in the gaps in our understanding of the $SFR-M*$ relation. Semi-resolved studies of clumpy galaxies show that stellar mass is closely tied to the number of UV-bright regions in a galaxy \citep{Martin2023} further motivating studies that use global and local galaxy properties in concert to disentangle star formation and feedback processes. Observations from JWST, where we can obtain excellent spectral resolution of very high redshift galaxies, are helping us to better understand the redshift evolution of galaxies through the correlation and can place constraints on outshining that cause biases to inferred stellar populations \citep[see][]{Harvey2025}. Large surveys made possible by Vera C. Rubin Observatory \citep{Ivezic2019}, Euclid \citep{Amendola2018Euclid} and Nancy Grace Roman Observatory \citep{Spergel2013, Spergel2015} will provide rich data sets with billions of galaxies to gain understanding of the faint and low mass limits. Among the many open questions about the $SFR-M_*$ correlation are: 1) How do galaxies evolve through $SFR-M_*$ space through their lifetimes? and 2) Does the correlation extend to low masses for regions within galaxies where stochastic processes regulate star formation on short timescales, and what is the associated slope and scatter?

The first question is an area where SFH reconstruction can aid in our understanding. For all but the lowest redshifts,  where individual stars within galaxies can be resolved, SFHs are reconstructed using spectral energy distribution (SED) fitting of spectra, IFU, or integrated photometry. Within the integrated light of the galaxy are encoded multiple episodes of star formation and the short and long timescale processes that have triggered them. While the SFR of a galaxy reflects only short timescale processes, SFH reconstruction infers the SFR of the galaxy back through cosmic time -- thus providing a window into how the galaxy was assembled through its stellar mass.   Studying SFH trajectories through the $SFR-M_*$ plane has been suggested as a novel way of tracing how galaxies evolve through $SFR-M_*$ space \citep[see][]{Iyer2018}.

The second question is an active area of study using both spatially resolved simulations and observations of galaxies. While scatter is expected to increase in the relation as a result of stochasticity and undersampling at small scales, the resolved star formation main sequence (rSFMS), as it is known, has been shown to hold to scales of a kiloparsec and below \citep{Feldmann2011, Kruijssen2014, CanoDiaz2016, AbdurroufAkiyama2017, Hsieh2017, Liu2018, Medling2018, Kruijssen2018, Ellison2020, Morselli2020, Pessa2021}. 
Several recent studies have sought to compare the SFMS and the rSFMS \citep{CanoDiaz2016, Hemmati2020, TrayfordSchaye2019, Jafariyazani2019, Abdurrouf2022, Yao2022}. These studies use observational tracers of properties such as H$\alpha$ and/or SED fitting. The SED fitting codes used in these prior studies assume parametric SFHs for their models such as a double power law \citep{Abdurrouf2022} or exponentially declining \citep{Hemmati2020, Jafariyazani2019, Yao2022}.
Pushing down to smaller physical scales allows for resolved studies to investigate the importance of the physical processes believed to drive galaxy evolution as a function of spatial scale. Short timescale processes may also have a larger impact locally.
Understanding the spatial distribution of \sigsfr and \sigmstar over time allows us to determine how much small scale processes drive global galaxy evolution. This is particularly the case when spatially resolved SFHs can be reconstructed so the evolution of these properties can be traced through cosmic time. Traditionally this work has been done using integral field spectroscopy (IFS) or integral field unit (IFU) surveys such as CALIFA \citep{Sanchez2012}, SAMI \citep{Croom12}, and MaNGA\citep{Bundy2015}, who have shown evidence of inside-out growth and downsizing \citep{Perez2013} existing at small scales, as well as $\Sigma_{SFR}$ correlating with $\Sigma_{M_*}$ over time and scaling with morphology \citep{Gonzalez2017}. These surveys offer excellent detail, but star formation information is hard to recover in the outskirts. Recent surveys have pushed the capacity of spatially resolved surveys into higher and higher redshifts to examine galaxy assembly in the early universe \citep{Rigby2025, Haryana2025, Abdurrouf2025, Tan2025, Sok2025a, Sok2025b}. These studies have revealed the complexity of galaxy formation at high redshifts, but a more comprehensive picture of how galaxies form in the early universe will become more clear when statistical samples become available from large surveys like the Legacy Survey of Space and Time (LSST) and Roman. SFHs of galaxies near enough to resolve individual stars have long been available through color-magnitude diagram (CMD) SFH reconstruction with impressive detail such as ANGST or its follow up PHAT \citep{dalcanton, Dalcanton2012}, though these surveys are limited statistically due to small sample size. Recent work has been able to utilize multiwavelength integrated photometry where neither IFS or IFU data are available \citep[see][]{Abdurrouf2022a}. Pixel SEDs created from integrated photometry have been shown to be consistent with results from IFU data up to high redshifts \citep[see][]{Gimenez-Artega2024}. With upcoming large photometric surveys providing a wealth of data, photometric resolved SED fitting promises a scalable solution when IFS and IFU data is not available that will allow for extensive study of resolved properties of galaxies at a variety of redshifts and environments to unearth a deeper picture of galaxy evolution.


The Cosmic Assembly Near-infrared Deep Extragalactic Legacy Survey (CANDELS) is an imaging survey carried out in the IR and optical using the Hubble Space Telescope with the WFC3 and ACS cameras respectively \citep{Grogin2011CANDELS, Koekemoer2011CANDELS}. CANDELS includes five multi-wavelength fields including the Great Observatories Origins Deep Survey (GOODS)-North and GOODS-South, the Extended Groth Strip (EGS), Cosmic Evolution Survey (COSMOS), and the Ultra-deep Survey (UDS). With deep observations allowing for detections of objects at z$\geq$8, CANDELS provides a means of probing galaxy evolution from cosmic dawn to the present day. To delve deeper into the science goals of CANDELS,  UVCANDELS \citep{Wang2024UVCANDELS,UVCANDELSDATA} imaged GOODS-N, GOODS-S, COSMOS, and EGS with deep, high-resolution WFC3 UVIS (F275W) and Blue (F435W) imaging that offer rest-frame UV coverage needed for  the study of star formation in galaxies at $z<0.3$.

For galaxies at $z<0.3$, the excellent resolution of these data  allows the reconstruction of SFHs for each region in these galaxies, enabling the generation of 2D property maps showing details of in-situ star formation. In addition, we can test the resolved $SFR-M_*$ relation with resolved photometric data and compare to the integrated scaling relation. 
Taking advantage of the excellent UV coverage for these galaxies we search for insights into star formation trends over the past Gyr and search for clues as to where we may expect to see the cessation of star formation in the future. We do this though the examination of the evolution of the resolved $SFR-M_*$ correlation and radial profiles of galaxy properties. We also compare the correlations for the sample to search for meaningful trends that may indicate inside-out or outside-in growth/quenching.

This paper is structured as follows: In Section~\ref{sec:UVData} we present the dataset and introduce the UVCANDELS Gold Sample. In Section~\ref{sec:Methods} we discuss our means for creating regions and our SFH reconstruction methods. In Section~\ref{sec:results} we present analysis for three example galaxies from the Gold Sample, first showing 2D maps of recovered properties for each region and then using resolved properties to test the resolved $\Sigma_{SFR}-\Sigma_{M_*}$ correlation  at the time of observation. In this section we also show key results from the radial profile analysis before summarizing our findings in Section~\ref{sec:summary}.

\section{Data}
\label{sec:UVData}
We analyze a series of postage stamps from the UVCANDELS catalogs\citep[see][]{Wang2024UVCANDELS, UVCANDELSDATA} located in the GOODS-N field \citep{Barro2019} and covering 10 bands --  WFC3 F275W, ACS F435W, ACS F606W, ACS F775W, ACS F814W, ACS F814W, ACS F8501LP, WCF3 F105W, WFC3 F125W, WFC3 F140W, and WFC3 F160W, ranging from 0.275 to 1.6 microns, covering UV-through-near-infrared light. Each postage stamp data cube has been PSF matched to the HST F160W band. These photometric catalogs are publicly available on the Barbara A. Mikulski Archive
for Space Telescopes.\footnote{{\href{https://doi.org/10.17909/8s31-f778}{https://doi.org/10.17909/8s31-f778}}}
\begin{figure}
    \centering
    \includegraphics[width=\linewidth]{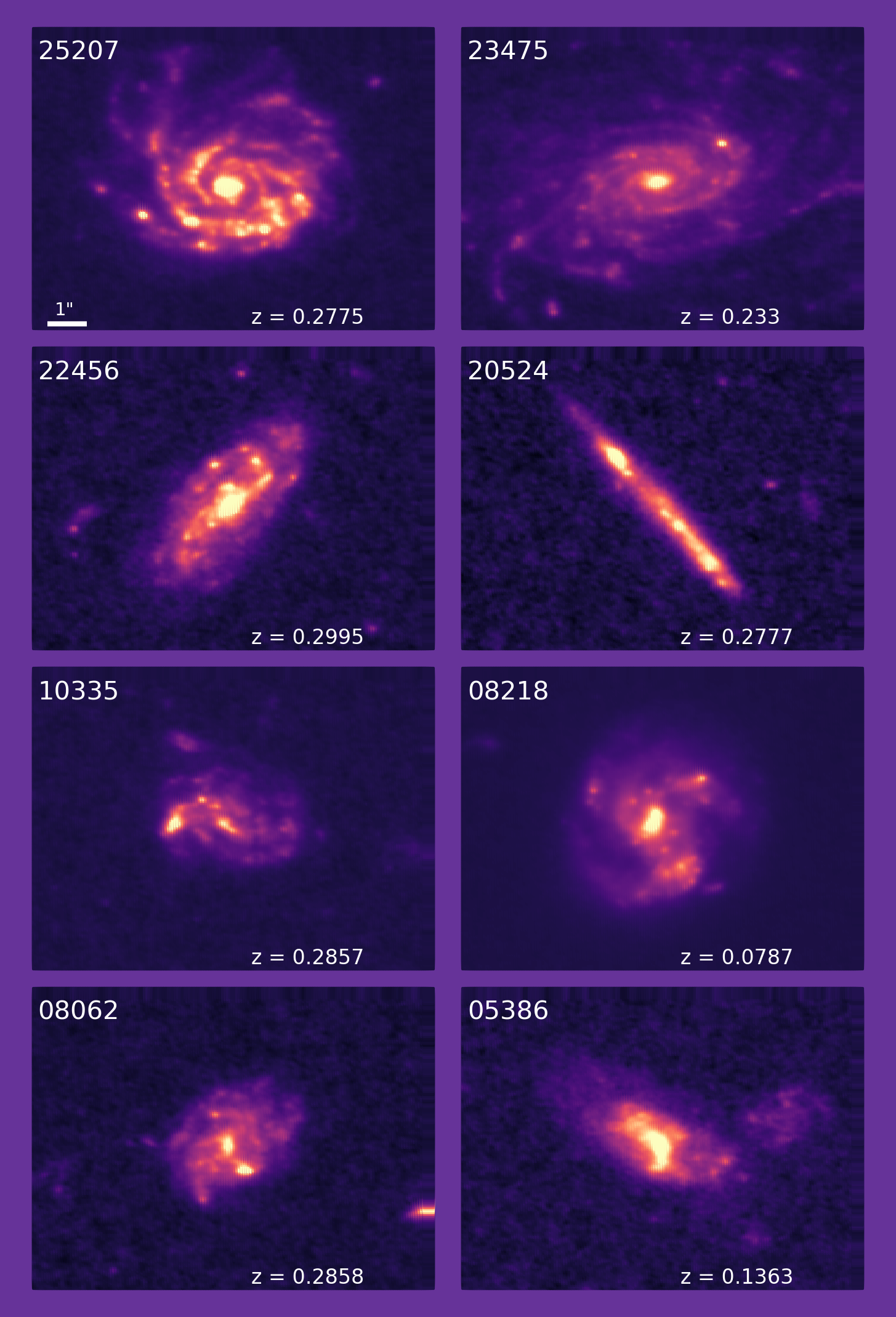}
    \caption{Eight galaxies of the 'Gold Sample' shown in the F435W band. These galaxies have a variety of inclinations and morphological features and have been chosen due to their high SNR in most bands and clear detection in both F275W and F435W.}
    \label{fig:multi_postage}
\end{figure}
From these postage stamps we select a ``Gold Sample'' of galaxies that are ideal for this analysis such that galaxies are through visual inspection 1) easily detected by eye in F275W and F435W, 2) are extended enough to span the majority of the 165 by 165 pixel postage stamp and show distinct features such as arms and clumps visible in multiple bands, and 3) are detected in all bands. This gold sample consists of 9 galaxies spanning a spectroscopic redshift range of 0.0787$<$z$<$0.9537 with the majority z$<$0.3, and consists mainly of face on galaxies supplemented with edge on disks and/or galaxies with interesting features or apparent tidal disturbances. In Table ~\ref{tab:Gold_Sample} we list the eight galaxies in the sample with their UVCANDELS ID, right ascension, declination, and redshift, and in Figure~\ref{fig:multi_postage} we show the postage stamps in F435W.  
\begin{table}[]
    \centering
    \begin{tabular}{||c|c|c|c||}
    \hline
    \hline
        ID & RA & Dec &   spec z  \\
        \hline
        \hline
         05386& 189.13528622 & 62.17706506 & $0.1363^{+0.0476}_{-0.0733} $\\
         \hline
         08062 & 189.05882385 & 62.19640837 & $0.2858^{+0.1262}_{-0.0032} $ \\
         \hline
         
         08218 & 189.15358828 & 62.19301442 & $0.0787^{+0.0732}_{-0.0397} $\\
         \hline
         
         10335 & 189.06628794 & 62.21040643 & $0.2857^{+0.1083}_{-0.0077} $ \\
         \hline
         20524 & 189.5382977 & 62.27724388 & $0.2777^{+0.1342}_{-0.0006} $ \\
         \hline
         
          22456 & 189.27697827 & 62.29142337 & $0.2995^{+0.0565}_{-0.0375} $ \\
         \hline
         23475 & 189.25633474 & 62.31184974 & $0.233^{+0.0200}_{-0.099}  $\\
         \hline
         25207 & 189.35635861 & 62.32816634 & $0.2775^{+0.0785}_{-0.0775}  $\\
         \hline
         \hline
    \end{tabular}
    \caption{The CANDELS IDs, right ascension, declination, and redshifts for the Gold Sample of galaxies from the GOODS-N postage stamps. }
    \label{tab:Gold_Sample}
\end{table}
 
\begin{figure}
    \centering
    \includegraphics[width=\linewidth]{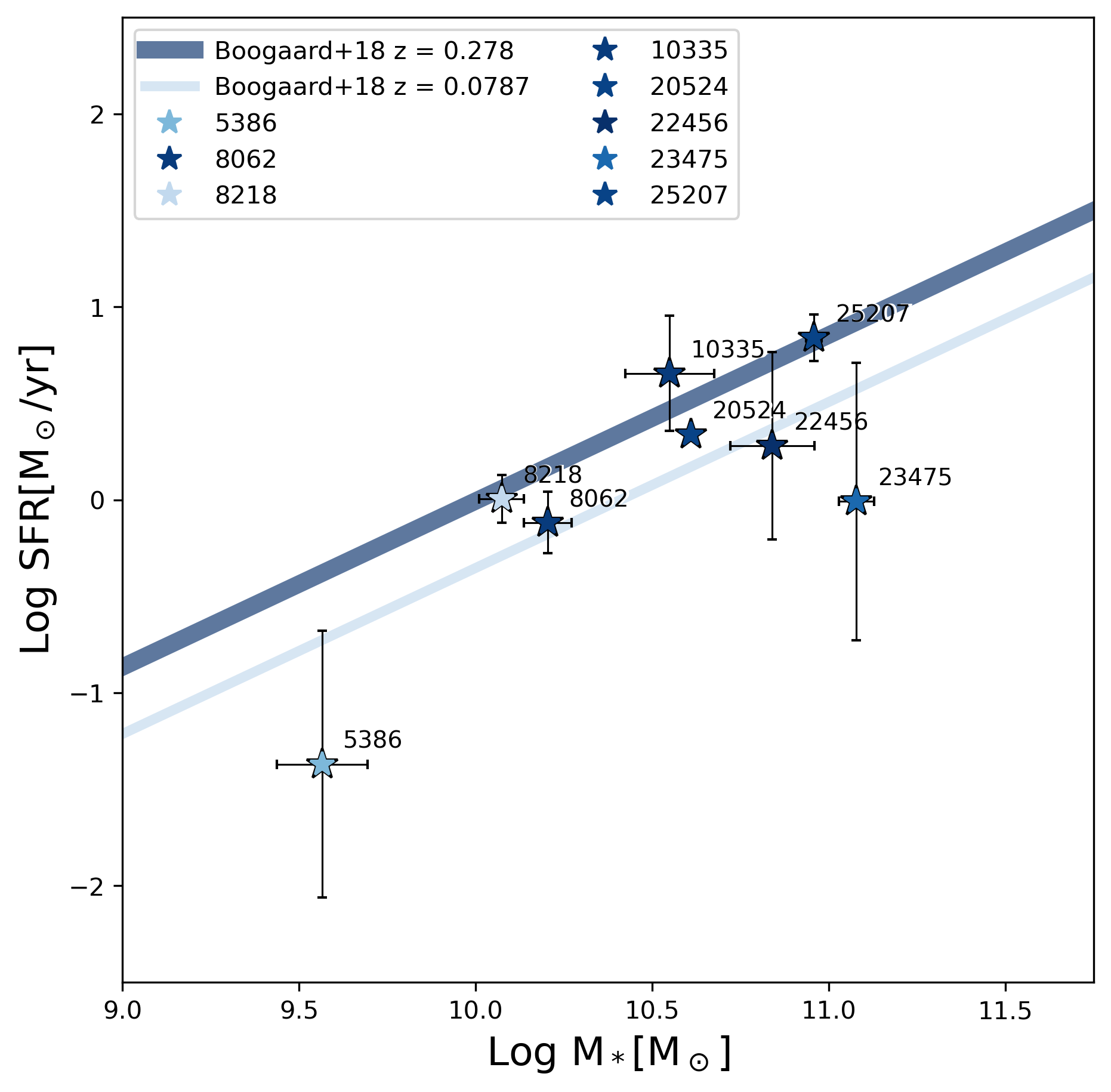}
    \caption{The log SFR and log stellar mass of the eight galaxies in our sample as calculated with the \texttt{densebasis} SED fitting code. Correlation lines from \cite{Boogaard2018} are shown for the median redshift for the sample and lowest redshift. All galaxies appear to lie on the correlation line with the exception of 05386 and 08062 which lie below their respective lines. }
    \label{fig:integrated_sfrmstar}
\end{figure}

We expect galaxies selected in part due to their features visible in the F275W band would naturally be star forming. To verify this, we plot the SFMS for global SFR and M$_*$ derived from integrated photometry and reported in \cite{Mehta2024}. We show this in Figure~\ref{fig:integrated_sfrmstar} with the correlation lines from \cite{Speagle2014}, \cite{Iyer2018}, and \cite{Boogaard2018}, with the latter correlation specifically relating to low z galaxies. Regardless of the correlation fit the galaxies appear to be star forming with the exception of UVCANDELS ID 05386 which lies below the correlation. 

\section{Methods}
\label{sec:Methods}
\subsection{Defining Regions}
We employ a Voronoi-like tessellation \citep{CappellariCopin2003, Wuyts2012, Chan2016, Hemmati2020, Fetherolf2020}. Starting from the brightest pixel in the F435W band, we grow regions by adding adjacent pixels until one of three criteria based on SNR, color, or distance from the central pixel is met. To avoid single-pixel regions in high-SNR areas, we require a minimum region size of four pixels in order to not sample below the limit of the F160W PSF for a combined pixel scale of 0.12". We adopt an SNR threshold of 10, a color-similarity $\chi^2$ threshold of 10, and a distance threshold of 10 pixels. Thus, regions around bright, high-SNR bulge pixels are typically limited by the SNR threshold. Otherwise, pixels are added until either the SNR threshold is exceeded or the $\chi^2$ threshold is met; if neither occurs, the region grows until it reaches the distance limit. This produces small, high-SNR regions near the galaxy center and large, low-SNR regions at the outskirts, yielding $\sim$40 to $>900$ regions per galaxy before cuts.

To avoid fitting empty areas of the postage stamp without explicitly imposing an SNR cut, we exploit the spatial continuity of the regions. We examine all regions in the segmentation map and, starting from the last region and moving backward, compute a rolling median SNR over 10\% of the total number of regions. When this median reaches an SNR of 30, we define that region as the maximum and retain all regions up to that index.

After cuts, each Gold Sample galaxy has $\sim$10–$\sim$800 regions with physical areas of $\sim$5–$\sim$30 kpc$^2$ or a diameter of up to $\sim$4 kpc.

\subsection{Spectral Energy Distribution Fitting}
 Although previous studies have used SED fitting for resolved observations, these works have not been able to benefit from the excellent UV imaging available from UVCANDELS which enables a fuller picture of recent star formation. Additionally, past studies have used SED fitting codes that assume a parametric SFH. Parametric SFHs are known to impose strong priors on physical parameters such as SFR and stellar mass, and these recovered properties in integrated observations are seen to bias scaling relations, including the $SFR-M_*$ correlation \citep{Carnall2019}. We used the \texttt{dense basis} \citep{iyer2017, Iyer2019} SED fitting code to fit the SEDs derived from the 10 bands in each of our regions defined by our segmentation map. \texttt{dense basis} is a fully nonparametric SFH reconstruction code based on Gaussian processes, which allows it to explore the entire functional space allows it to be flexible enough to recover reliable SFHs and has been validated against galaxy formation simulations as well as low redshift dwarf galaxies \citep{Olsen2021}. 
 \texttt{dense basis} has been validated against SFHs reconstructed from color-magnitude diagrams \citep{Olsen2021} and can more reliably capture stochasticity in star formation on smaller physical scales than parametric models. Unlike other nonparametric SED fitting codes that explore the parameter space using techniques such as an MCMC, \texttt{dense basis} cuts computational costs by generating an atlas of models that can be fit numerous times making it fast without sacrificing reliability. While reconstruction of $\sim$ 100 SFHs for a single galaxy would be computationally costly for many codes, \texttt{dense basis} can reconstruct an SFH in $<$1 second per SED, and all SFHs from a single galaxy can be reconstructed in a single run due to them sharing a redshift. Due to the size and estimated mass of each of the regions, we started with the \texttt{dense basis} priors used in  \cite{Olsen2021} and treated individual regions as if they were low mass galaxies. We changed the lower bound of the stellar mass prior to $10^3$M$_\odot$ to better represent the lowest mass regions. Since all galaxies in the sample have spectroscopic redshifts, we set a narrow prior around redshift so that individual regions reflect the global spectroscopic redshift for the galaxy. Using \texttt{dense basis} on each region returns the SFH as a tuple of SFR, stellar mass, and $t_X$ parameters. The $t_X$ parameters are lookback times at which the region has formed a specific fraction of its stellar mass. We chose to use three $t_X$ parameters, making our $t_X$ parameters $t_{25}$,$t_{50}$, and $t_{75}$, or the lookback times at which a region formed the first 25\%, 50\%, and 75\% of its stellar mass. 

Performing this analysis on each region allowed us not only to recover a full SFH for each region, but also to create 2D maps for properties such as $\Sigma_{SFR}$, $\Sigma_{M_*}$, $t_{50}$ and $A_V$. For this the properties were recovered for each SED and then divided by the area of each region before plotting. 

\subsection{Resolved Properties of Regions in Galaxies}

\begin{figure*}
    \centering
    \includegraphics[width=0.95\linewidth]{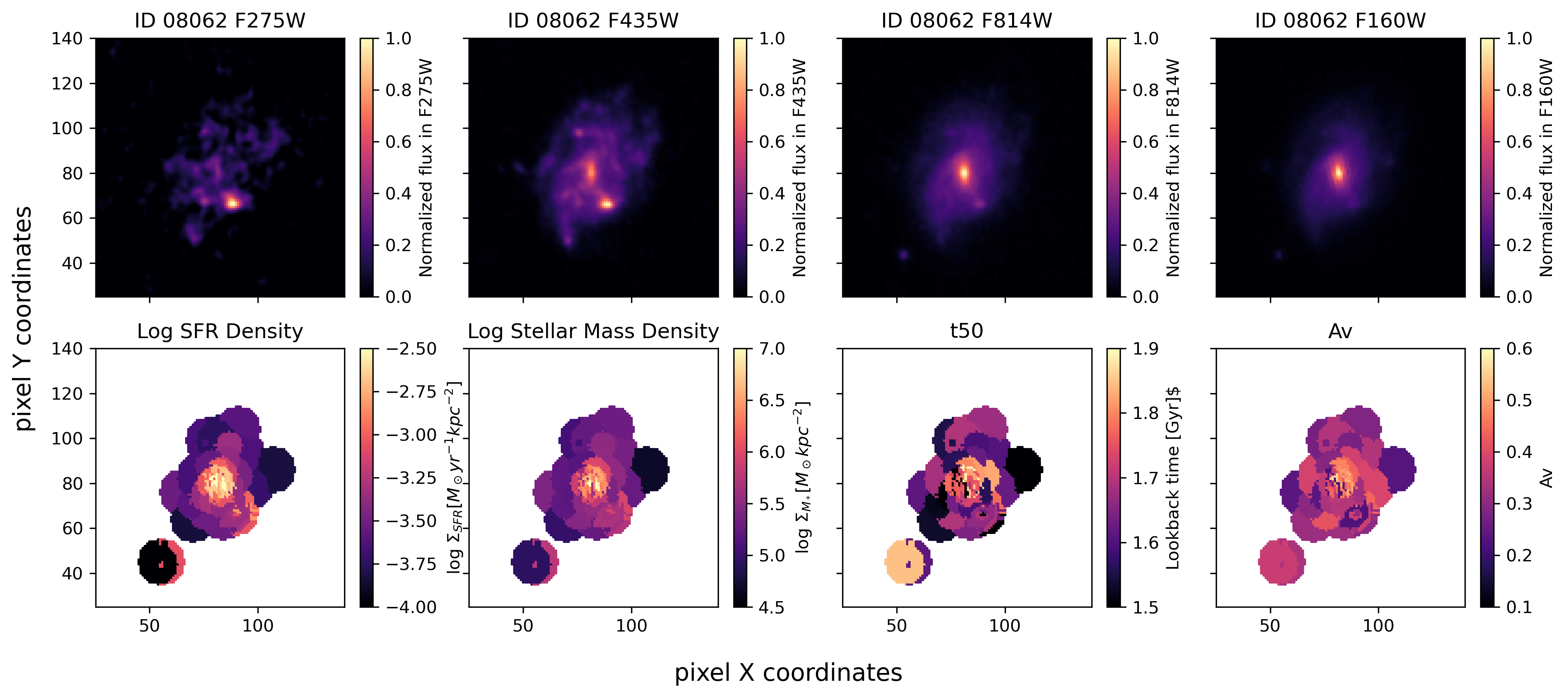}
    \caption{UVCANDELS 08062 has a bright off-center feature most clearly seen in F275W and F435W, with F435W also revealing internal structure. The segmentation map shows peak SFR density at the center, with elevated SFR density at the bright feature just right of center. The SFR and stellar mass density maps both show an arc of enhanced density above this feature. The $t_{50}$ map indicates the bright clump is younger than the rest of the galaxy, with additional young regions toward the bottom and a smaller knot visible only in F435W in the upper right. $A_V$ is fairly uniform, except for higher values on the bulge.}
    \label{fig:multi_08062}
\end{figure*}

\begin{figure*}
    \centering
    \includegraphics[width=0.95\linewidth]{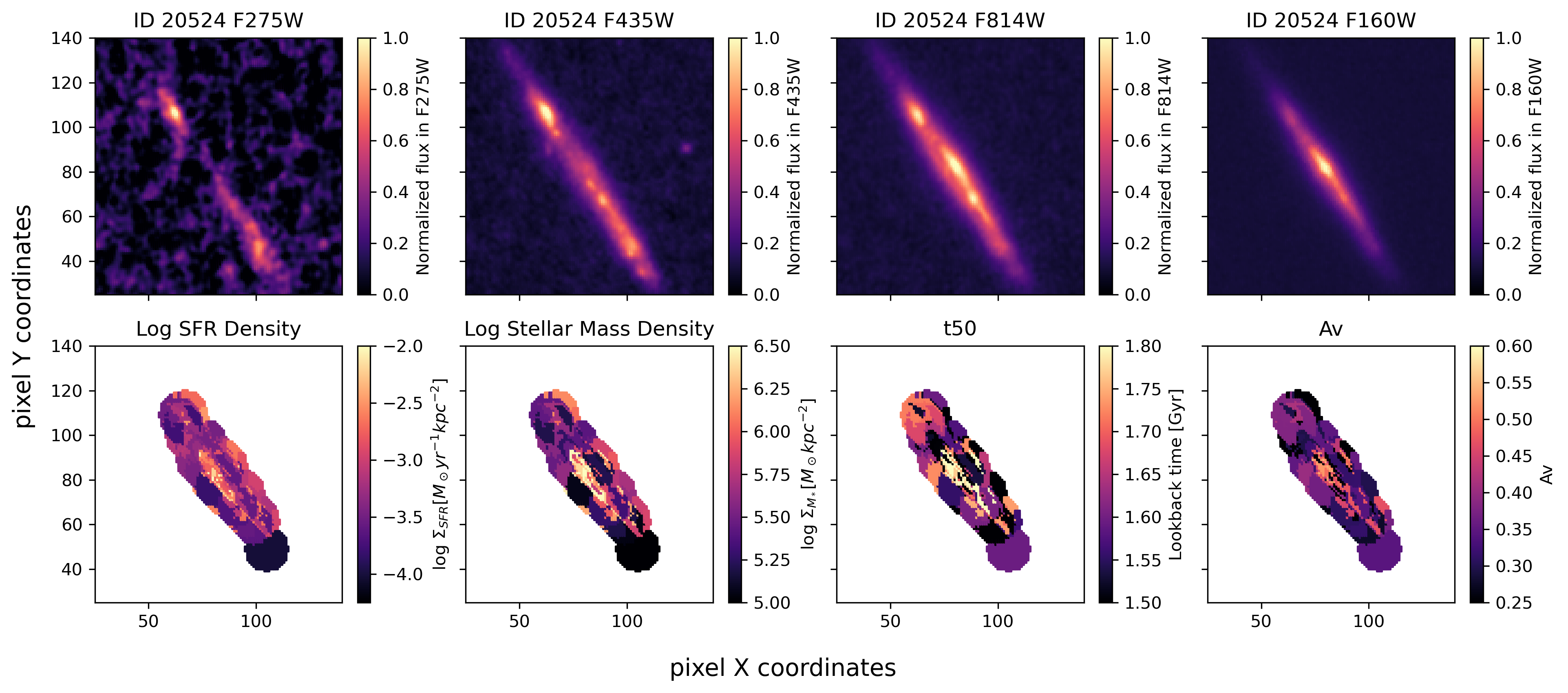}
    \caption{Postage stamps of the edge-on galaxy UVCANDELS 20524 show that while the F160W image is bulge-dominated, the F275W image is bright mainly in the outskirts—especially the upper left—rather than the center. This is consistent with dust obscuration, as supported by the $A_V$ map. However, the UV-bright region in the lower right does not show elevated properties, including $A_V$, ruling out dust obscuration there. The SFR and stellar mass density maps show enhanced values along the top edge of the disk. The $t_{50}$ map reveals both older- and younger-than-average regions in this area. Overall, these features may result from projection effects in an edge-on galaxy.}
    \label{fig:multi_20524}
\end{figure*}

\begin{figure*}
    \centering 
    \includegraphics[width=0.95\linewidth]{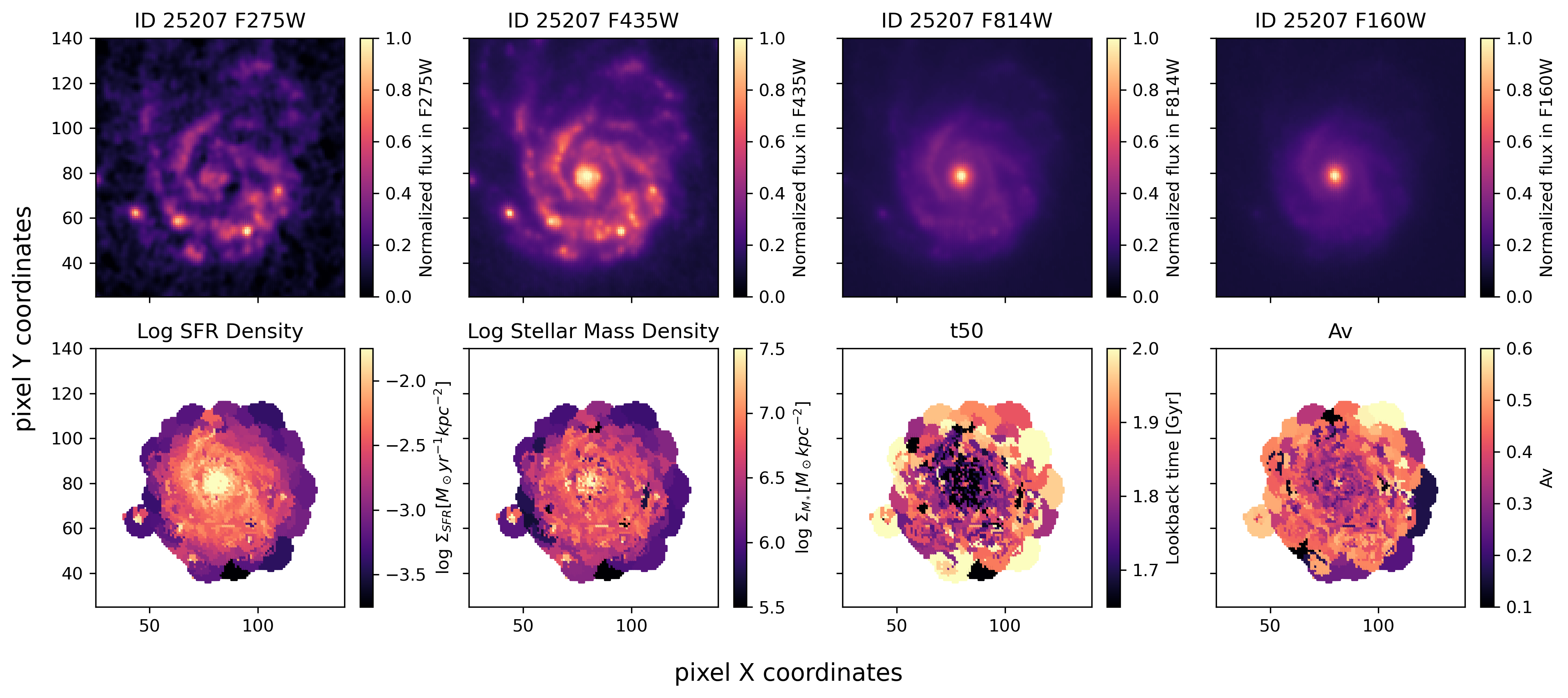}
    \caption{Postage stamps for the face-on spiral UVCANDELS 25207 show in F435W a bright bulge and clear spiral arms. In F275W the spiral arms are also clear, with bright knots along the lower arm and a particularly bright knot at the apparent end of an arm on the far left. This knot, and several lower-image clumps, appear in the SFR density map, though these knots are small and hard to distinguish in all four property maps. Spiral arms are faintly visible in SFR density, but stellar mass density and $A_V$ show no strong patterns. The $t_{50}$ map, however, shows younger regions tracing the spiral arms and central bulge.}
    \label{fig:multi_25207}
\end{figure*}

After defining regions, we fit their SEDs with the \texttt{dense basis} code. Figures ~\ref{fig:multi_08062}–\ref{fig:multi_25207} show, in the top row, 2D intensity maps of each galaxy in normalized flux in the F275W, F435W, F814W, and F160W bands. The second row shows 2D maps of regions colored by Log $\Sigma_{SFR}$, Log $\Sigma_{M_*}$, $t_{50}$ (the time when half the stellar mass formed), and $A_V$, all reconstructed from the SED fits at the time of observation. Our galaxies exhibit features similar to the bright clumps in star-forming systems \citep[e.g.,][]{Martin2023, Giunchi2025}, though these appear fainter in the property maps because the regions often span larger areas than the clumps.

We present these property maps for all eight galaxies to qualitatively explore their formation histories via spatial variations in these quantities. The multi-band images show which structures are visible in each band and, by comparison with the property maps, which are captured by the tessellation. In this Section we show three examples—UVCANDELS 08062, 20524, and 25207; the remaining six are in Appendix~\ref{appendix:propmaps}.

Figure~\ref{fig:multi_08062} shows UVCANDELS ID 08062 at $z = 0.2858$. The postage stamps and segmentation maps are colored by region properties. In the top row, a bright F275W knot appears slightly offset from the galactic center seen in the other three bands. This knot is also visible in F435W and faintly in F814W, but not clearly in F160W. The bottom row shows segmentation maps defining 55 regions. The bottom-left panel shows the highest \sigsfr at the galaxy center, at the UV-bright region, and in a crescent-shaped outer area around the UV-bright knot. Although this knot is not strongly enhanced in \sigmstar relative to its surroundings, the crescent is visible. The $t_{50}$ map indicates that the UV-bright region is younger than much of the galaxy and that a bright F435W spot above the center is also younger than its surroundings. The $A_V$ map follows the galaxy’s morphology but shows little variation.

Figure~\ref{fig:multi_20524} shows the edge-on galaxy UVCANDELS ID 20524. In F160W it appears relatively smooth, brightest near the bulge. Structure is clearer in F814W, and F435W reveals a bright upper disc region, fainter emission elsewhere, and several bright spots in the plane. F275W also highlights the bright upper disc and brighter areas near the lower disc, consistent with F435W. Before our cutoff, UVCANDELS ID 20524 contained 60 regions. \sigsfr is elevated near the center and along the plane. The upper bright region has lower overall \sigsfr, though a small enhancement coincides with the brightest F275W pixels. Segmentation maps of \sigsfr and \sigmstar show depressed values immediately around the center on both sides of the disc, with enhanced \sigsfr and \sigmstar beyond this on the galaxy’s right. This feature is not explained by the image stamps: $t_{50}$ there is similar to the rest of the galaxy, and $A_V$ is fairly uniform except for higher dust in the plane. A possible caveat is that since photons from stars located deeper within the galaxy must travel through more dust than those from stars near the edge, there is a possibility of strong differential dust attenuation along the plane, with the stellar populations closer to the center experiencing higher optical depths and a flatter attenuation slope \citep{Chevallard2016, Lower2022ApJ, Nagaraj2022ApJ, Sommovigo2025}. While modeling this differential dust attenuation is outside the scope of this paper, we note that 
if \texttt{dense basis} underestimates the $A_V$ in areas of higher attenuation caused by the edge on nature of the galaxy, this could lead to an underestimate of the stellar mass for those regions.

Figure~\ref{fig:multi_25207} shows UV CANDELS ID 25207, a face-on spiral with 550 regions above the SNR cutoff. In F160W it has a prominent bulge with clear spiral arms. F275W and F435W reveal star-forming knots along the lower arms; these knots are faint in F814W and not clearly seen in F160W. The \sigsfr map highlights these clumps with elevated \sigsfr relative to their surroundings. The spiral arms also show slightly enhanced \sigsfr, most clearly in the arm curving toward the top of the frame. We also see elevated \sigmstar in the bright knots, though the arm regions partly obscure the \sigsfr structure. The $t_{50}$ map shows younger regions along the arms, while $A_V$ shows no strong trends.

Overall, the \sigsfr and \sigmstar maps best reproduce the features seen in F435W and F275W. Some features are subtle in the property maps, while others are very clear. The \sigsfr and \sigmstar maps are physically consistent with the galaxies, even when region definitions obscure structures such as spiral arms. The $t_{50}$ and $A_V$ maps are less consistent: some show only weak trends (as in 05386), whereas in other cases $t_{50}$ or $A_V$ trace galaxy features more clearly than the other properties. Although some $A_V$ maps appear disordered, they agree with higher-redshift results \citep{Kim2017, Kim2019}.

\section{Results}
\label{sec:results}

 We compare the resolved $\Sigma_{SFR}$ and $\Sigma_{M_{*}}$ of regions within each galaxy to the integrated values at the time of observation and 1 Gyr lookback time. The choice of 1 Gyr comes mainly from our concerns about stellar mixing. Since the crossing time for a galaxies in our sample galaxy is anywhere between 300 Myr and 1 Gyr we place a constraint of 1 Gyr lookback time as the boundary at which we expect our SFHs to still be reliable. 1 Gyr lookback time in SFHs also corresponds to the timescale at which A stars are no longer present, making the SEDs less sensitive \citep{Dressler2018, Peterkin2020}. We analyze how the \sigsfr–\sigmstar correlation varies from galaxy to galaxy and how it evolves. Specifically, we recover the slope of the resolved \sigsfr–\sigmstar relation for all regions in our sample and compare the slopes for individual galaxies with that of the combined ensemble. We further examine differences in normalization between individual-galaxy and ensemble relations, and any dependence on the scatter width for each galaxy. 

The property maps, projected radial profiles, and resolved $\Sigma_{SFR}$–$\Sigma_{M_{*}}$ correlations shown here are for three example galaxies; results for the remaining galaxies are provided in the Appendices.

\subsection{Spatially resolved evolutionary tracks in the SFR-$M_{*}$ relation}
\label{sec:res_sfrmstar}
Resolved scaling relations indicate the spatial scales where different physical processes dominate. 
Like the integrated $SFR-M_*$ correlation, the resolved relation evolves with redshift, and simulations by \cite{TrayfordSchaye2019} show that its slope increases with redshift.
The resolved relation traces small-scale, internal evolutionary processes, such as inside-out or outside-in growth, which must still match the global trends in the integrated relation for galaxy evolution theories to remain valid. Interpreting resolved relations alongside integrated ones connects processes across orders of magnitude and reveals each galaxy’s distinct evolutionary path.
\begin{figure*}[h!]
    \centering
    \includegraphics[width=0.95\linewidth]{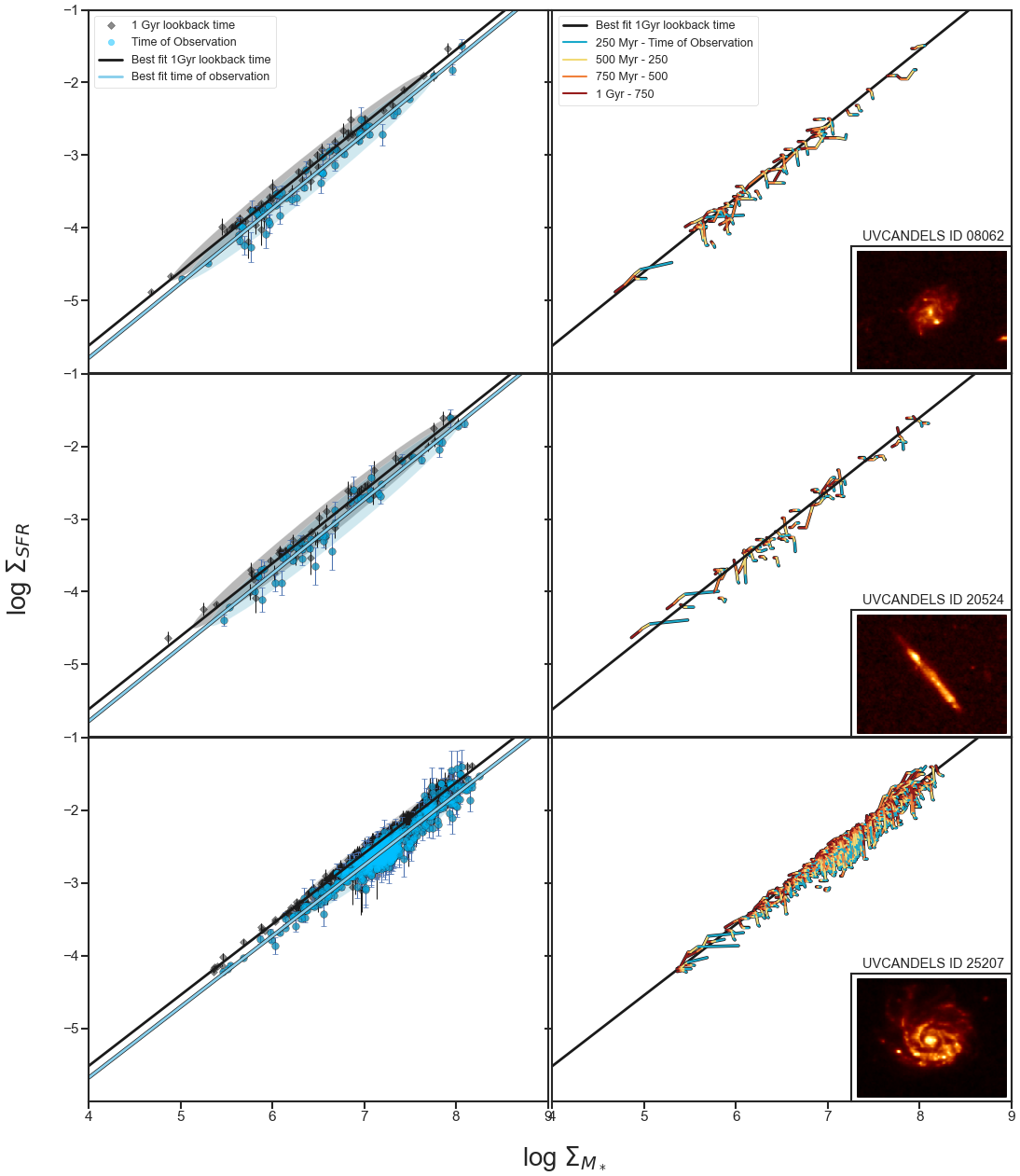}
    \caption{Evolution of the \sigsfr–\sigmstar correlation over the past Gyr for three galaxies (UVCANDELS IDs 08062, 20524, 25207). \sigsfr and \sigmstar are measured at the time of observation (cyan) and 1 Gyr lookback time (black), with correlation lines fit in the left panel for each galaxy. We also compute properties at 250, 500, and 750 Myr lookback time and trace the evolution of each region in \sigsfr–\sigmstar space in the right panel. Inset F435W images of each galaxy are shown for reference. Galaxies 08062 and 20524, which have fewer regions than 25207, exhibit more stochastic evolutionary trajectories.}
    \label{fig:sfr_mstar_example}
\end{figure*}

Although the $SFR-M_*$ scaling relation may hold down to kpc scales, the slopes reported for the resolved $\Sigma_{SFR}-\Sigma_{M_*}$ relation range widely from 0.6 to 1.4 \citep{CanoDiaz2016, AbdurroufAkiyama2017, Hsieh2017, Pessa2021, Yao2022}. Even within the same redshift range, studies differ due to fitting methods, selection effects, and how SFR and $M_*$ are measured. For example, \cite{Jafariyazani2019} applied different fitting techniques to CANDELS + MUSE data at $0.1<z<0.4$ and found a slope of 0.78 using SED-derived $\Sigma_{SFR}$, but 0.43 using H$_\alpha$. Although methodological systematics affecting the resolved $\Sigma_{M_*}$–$\Sigma_{SFR}$ relation are recognized, less is known about how slope and normalization vary between galaxies. Using the resolved regions of these eight galaxies, we test whether galaxy structure influences this correlation and whether the qualitative trends indicate inside-out or outside-in quenching. With reconstructed SFHs for each region, we can also track the evolution of the relation back to when stellar mixing becomes dominant.

From the best-fit SFHs, we measure \sigsfr and \sigmstar at the time of observation and at lookback times of 250 Myr, 500 Myr, 750 Myr, and 1 Gyr. We plot the correlation at 1 Gyr lookback time and at the time of observation, and connect each region across the intermediate epochs to trace its evolution over the past Gyr.
Figure~\ref{fig:sfr_mstar_example} shows the correlation between \sigsfr and \sigmstar for regions in each galaxy at lookback times of 1 Gyr, 750 Myr, 500 Myr, 250 Myr, and at the time of observation. The left panels show the regions and best-fit lines for 1 Gyr (black) and the time of observation (cyan), with 2$\sigma$ confidence ellipses; the right panels show the evolution of each region across these epochs. Additional examples are provided in Appendix~\ref{appendix:sfrmstar}. 

The example galaxies exhibit diverse evolutionary tracks, especially for regions in 08062 and 20524. Galaxy 25207 shows clear tracks indicating an overall decrease in \sigsfr for most regions. In all three galaxies, the correlation line at the time of observation has a lower normalization than at 1 Gyr, as seen in the left-hand panels of Figure~\ref{fig:sfr_mstar_example}.  

UVCANDELS ID 08062 spans a wide range of \sigsfr and \sigmstar, with most regions in the range of $5.77 \leq$ \sigmstar $\leq 8$. Many regions at the time of observation lie below the 1 Gyr correlation line, and some even below the more recent relation, as indicated by the black diamonds. This implies diverse SFHs but an overall decline in Log sSFR with time. Most regions have increased \sigsfr, but \sigmstar has grown enough to move them below the 1 Gyr trend. Thus, while the global 1 Gyr evolution shows a net decrease in \sigsfr, individual regions follow varied evolutionary paths to reach this outcome.
UVCANDELS ID 20524 is an edge-on galaxy with 60 regions, leading to a relatively sparse \sigsfr–\sigmstar diagram. The evolutionary tracks show highly diverse SFHs: some regions decrease in \sigsfr over ~1 Gyr, while others undergo strong increases so that regions at 1 Gyr lookback time lie closer to the time-of-observation correlation.Most regions show only minor \sigsfr increases, insufficient to keep them on the 1 Gyr correlation. As in other galaxies, 20524 thus has a correlation at the time of observation with a lower normalization than 1 Gyr ago. Its slope is also slightly steeper but very similar to that at 1 Gyr. (see Table~\ref{tab:slopes_and_norms}).

For UVCANDELS ID 25207, the correlation at the time of observation again has a lower normalization than at 1 Gyr lookback time. Here, the slope change is not visually apparent, and there is no clear trend of stronger quiescence in smaller, outer regions compared to the more massive central regions, though many regions lie between $7 \leq$ \sigmstar $\leq 7.25$.

We expect a different normalization for resolved galaxies even if their integrated values lie on the integrated $SFR-M_*$ correlation, as dividing any region by area as we have done will move it down and to the left of the correlation with a slope of one. This shift would steepen the slope and shift the normalization down relative to the integrated correlation. Since the resolved correlation accounts for area, a slope of 1 would imply that each region had a similar SFH. In general, the steeper slope has implications for the growth of these galaxies.
\cite{Yao2022} reports a slope of 0.771 with an intercept of -7.812. While broadly consistent with the \cite{Yao2022} result and the SED fitting result from \cite{Jafariyazani2019} our fit for each of our galaxies shows a steeper slope both at the time of observation and at 1 Gyr lookback time, as well as a lower normalization.

 Within our sample, we find a variety of slopes, each with different normalizations, as reported in Table~\ref{tab:slopes_and_norms}. The correlations for these galaxies individually give us clues to that galaxy's growth, while common features hint at broader trends in galaxy evolution. 

\renewcommand\theadalign{c}
 \begin{table*}[]
    \centering
   \begin{tabular}{|l|c|c|c|c|}
     \hline
    \hline
    UVCANDELS ID & Slope  & Slope  & Normalization & Normalization   \\
    {} & Time Of Observation & 1 Gyr Lookback Time& Time Of Observation & 1 Gyr Lookback Time \\
    \hline
     \hline
     05386 & $1.0911 \pm 0.0211$ & $0.9968 \pm 0.0118$ & $-10.4093\pm 0.1651$ & $-9.5991 \pm 0.0711$\\
     \hline
     08062 & $1.0196 \pm 0.0210$ & $1.0172 \pm 0.0106$ &  $-9.8785 \pm 0.1640$& $-9.6891 \pm 0.0675 $\\
     \hline
     08218 & $1.0184 \pm 0.0027$ & $0.9744 \pm 0.0066$ &  $-9.8868 \pm 0.0202$& $ -9.3981 \pm 0.0137$\\
     \hline
     10335 & $0.9935 \pm 0.0146$ & $0.9702 \pm 0.0113$ &  $-9.6848 \pm 0.1140$& $-9.3926 \pm 0.0451$\\
     \hline
     20524 & $1.0084 \pm 0.0224$ & $0.9985 \pm 0.0037$ &  $-9.8081 \pm 0.1746$&  $-9.6049 \pm 0.0743 $\\
     \hline
     22456 & $0.9545 \pm 0.0100$ & $0.9902 \pm 0.0034$ & $-9.3827 \pm 0.0791$& $-9.5015 \pm 0.0268$ \\
     \hline
     23475 & $1.0284 \pm 0.0025$ & $0.9759 \pm 0.0034$ & $-9.9451 \pm 0.0189$ & $-9.3972 \pm 0.0247$\\
     \hline
     25207 & $0.9300 \pm 0.0099$ & $0.9729 \pm 0.0043$ &  $-9.2081 \pm 0.0790$& $-9.4060 \pm 0.0312$\\
     \hline
     \hline
    \end{tabular}
    \caption{The CANDELS IDs, correlation slope at time of observation and at 1 Gyr lookback time, and the correlation normalization at the time of observation and 1 Gyr lookback time. }
    \label{tab:slopes_and_norms}
\end{table*}

\subsection{Examining trends between the correlation lines of galaxies}
 \begin{figure*}
    \centering
    \includegraphics[width=0.9 \linewidth]{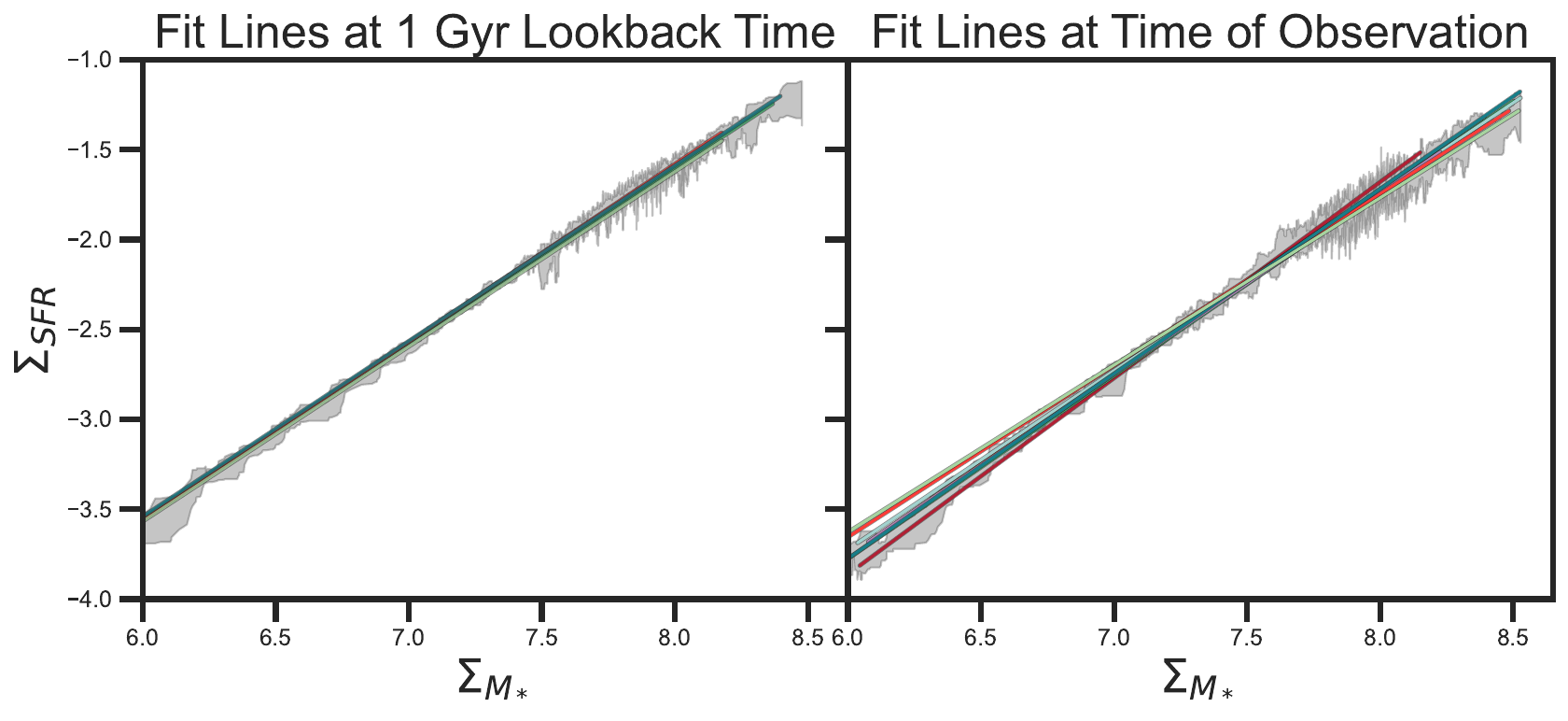}
    \caption{\sigsfr–\sigmstar correlation fit lines for all galaxies are shown on the Left at 1 Gyr lookback time and on the Right at the time of observation. The gray shading shows the rolling $1 \sigma$ confidence interval for all regions of all galaxies. At 1 Gyr lookback time, all galaxies lie on the same $SFR-M_*$ correlation, but the differing slopes and normalizations at the time of observation suggest that galaxies have since begun quenching in different ways.}
    \label{fig:fit_lines}
\end{figure*}

\begin{figure*}
    \centering
    \includegraphics[width = 0.98\linewidth]{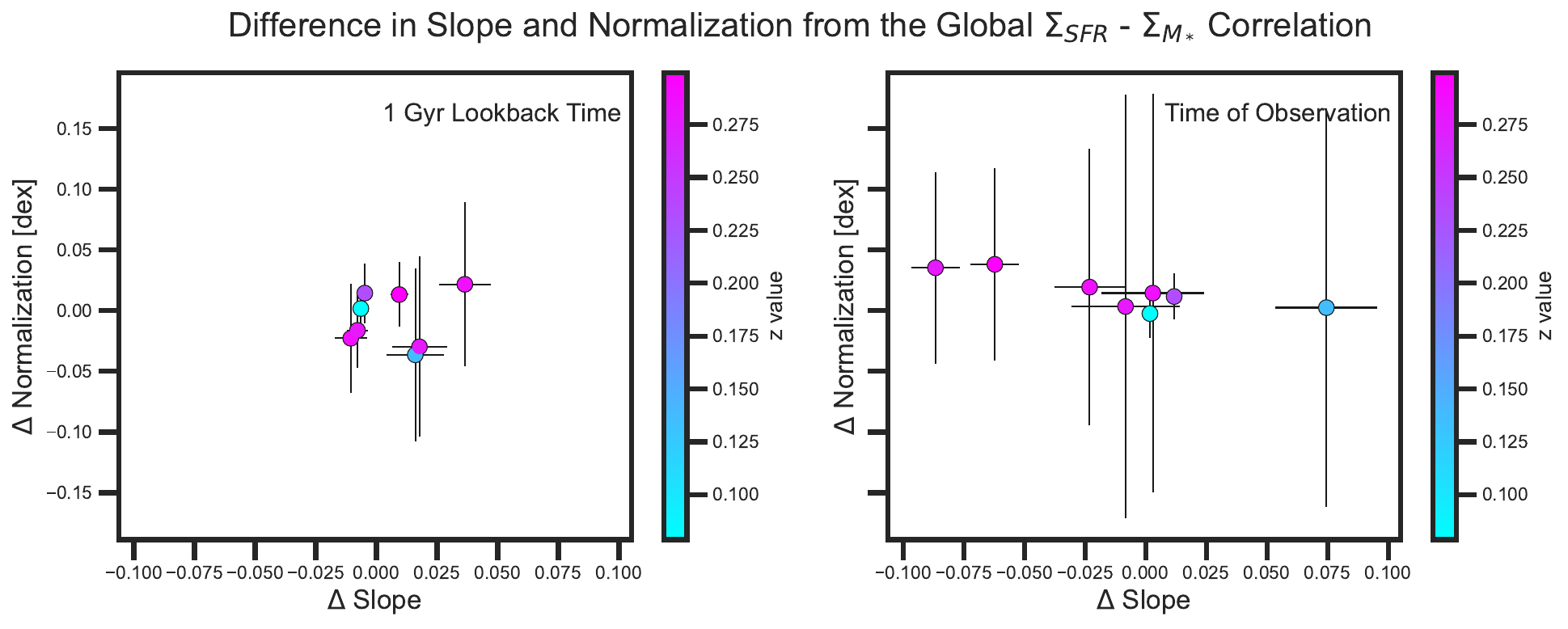}
    \caption{The two panels show how the slope, normalization, and redshift vary between each galaxy’s regional \sigmstar-\sigsfr correlation and the global correlation line centered at \sigmstar = 7.25 [$M_\odot$ $kpc^{-2}$]. The left and right panels show 1 Gyr lookback time and the time of observation, respectively, colored by redshift, with outlier 02042 marked by a black star. At 1 Gyr, slopes are nearly uniform across galaxies, though normalizations differ. At the time of observation, slopes vary while normalizations are similar. Redshift shows no clear correlation with either slope or normalization. }
    \label{fig:slope_offsetz}
\end{figure*}
 \begin{figure*}
    \centering
    \includegraphics[width=0.98\linewidth]{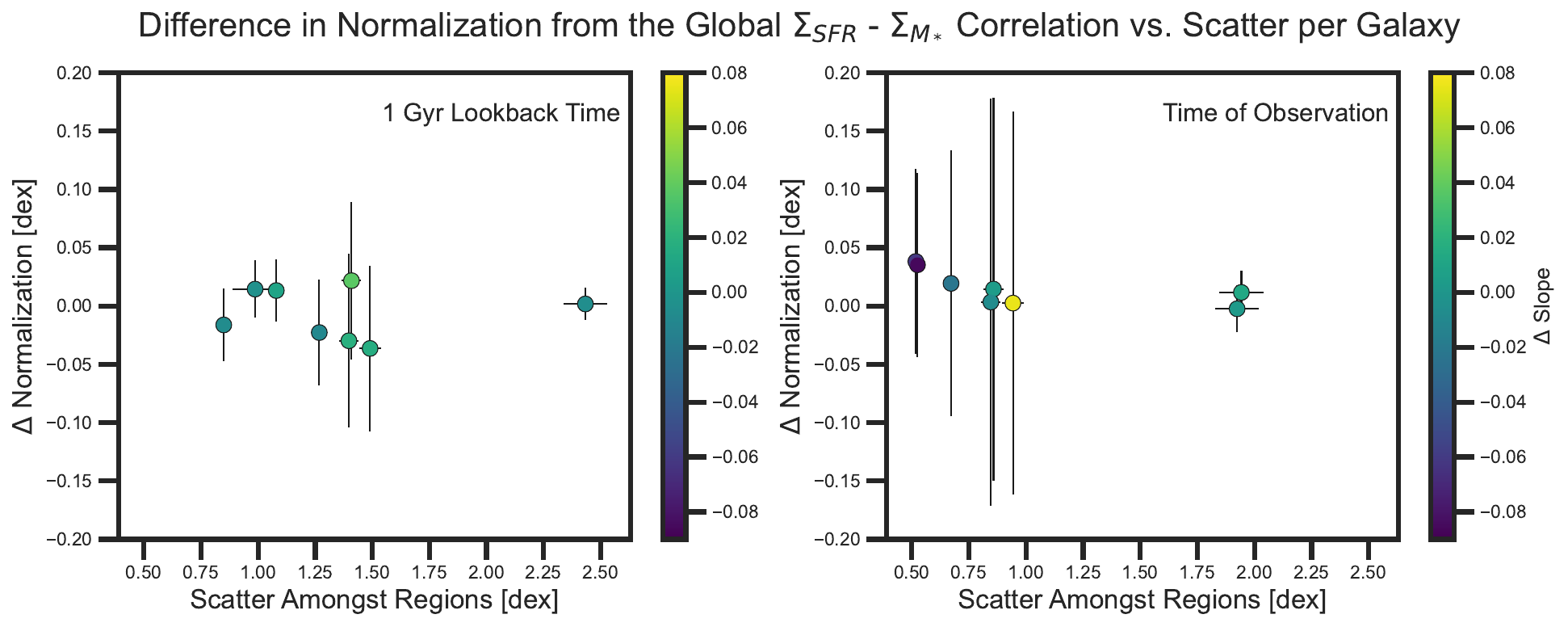}
    \caption{The two panels show how the slope, normalization, and scatter of each galaxy’s regional \sigmstar-\sigsfr correlation differ from the global relation, which is centered on \sigmstar = 7.25 [$M_\odot$ $kpc^{-2}$]. The left panel shows 1 Gyr lookback time and the right panel the time of observation. At 1 Gyr, slopes are nearly identical across galaxies, but normalizations differ somewhat. At the time of observation, slopes vary while normalizations are similar. The scatter in each galaxy’s fit does not correlate with either slope or normalization. }
    \label{fig:slope_offset}
\end{figure*}

The evolving \sigsfr–\sigmstar correlation over 1 Gyr traces how star formation changes across each galaxy. \cite{Mehta2024} show these galaxies lie on the integrated $SFR$–$M_*$ relation for their redshifts. To compare resolved behavior with the galaxy population, we first fit all regions from all eight galaxies together, then compare each galaxy to this global relation.

Figure~\ref{fig:fit_lines} shows the fits for all galaxies at 1 Gyr lookback time and at time of observation (left to right). The gray shaded regions give a rolling $1\sigma$ confidence interval for all regions. At 1 Gyr lookback time, the \sigsfr–\sigmstar relations agree closely.

At time of observation, the slopes and normalizations diverge slightly and appear to pivot around \sigsfr $\approx -2.5$ [$M_\odot$ yr$^{-1}$ kpc$^{-2}$] and \sigmstar $\approx 7.25$ [$M_\odot$ kpc$^{-2}$], where the scatter is minimal. Although this pivot is absent at 1 Gyr lookback time, the scatter in this same region of parameter space is also reduced there.

For a population with initially consistent \sigsfr–\sigmstar correlations, some spread in slope and normalization over 1 Gyr is expected. What is striking is the apparent pivot point in \sigsfr–\sigmstar space about which this spread occurs, and around which the global scatter is strongly suppressed, as indicated by the gray regions in Figure~\ref{fig:fit_lines}. The spread in slopes at time of observation is still modest, but at 1 Gyr lookback time all galaxies share an almost identical slope, implying more uniform \sigsfr at both high and low \sigmstar. This similarity is not entirely surprising, but may also reflect stellar mixing that produces similar SFHs in neighboring regions at that epoch, a possibility that warrants further study.

In Figure~\ref{fig:slope_offsetz} we re-normalize the relation for each galaxy, and the combined correlation, by centering on the apparent pivot point \sigmstar = 7.25[$M_\odot$ $kpc^{-2}$]. We then measure the differences in slope and normalization between each galaxy’s correlation and the global fit, shown for 1 Gyr lookback time (left) and the time of observation (right), with points colored by redshift. Redshift does not correlate with these differences. Instead, we find a larger spread in normalization at 1 Gyr lookback time and a larger spread in slope at the time of observation.  

Thus, re-centering both epochs on \sigmstar $\approx 7.25$[$M_\odot$ $kpc^{-2}$] highlights that the normalization spread present at 1 Gyr lookback time disappears at the time of observation, where the variation is instead dominated by slope.  

To test whether these trends might correlate with the width of each galaxy’s confidence interval, we repeat the analysis, now coloring points by the change in slope (Figure~\ref{fig:slope_offsetz}) and plotting the change in normalization against the within-galaxy scatter. We find no correlation between the confidence-interval width and the change in either slope or normalization.

Figures ~\ref{fig:fit_lines}, ~\ref{fig:slope_offsetz}, and ~\ref{fig:slope_offset} highlight a common pivot point and mass across all galaxies. The mass \sigmstar $\approx$ 7.25[$M_\odot$ $kpc^{-2}$] suggests that processes on this scale may regulate star formation in MW-type disc galaxies like those studied here.

Around this pivot point we identify two populations: one with excess star formation in low-mass regions and reduced activity in the dense center, and another showing the opposite. These patterns point to inside-out and outside-in quenching mechanisms—such as AGN, SN feedback, or varying relic inflows from the outer gas disk—affecting each galaxy differently.

\subsection{Radial profiles of properties across time}

Small changes over time in the slope and normalization of the \sigsfr-\sigmstar correlation can suggest spatially dependent galaxy evolution, but to better probe internal structure we examine projected radial profiles of \sigmstar, \sigsfr, $A_V$, and the fractional mass formation times $t_{25}$, $t_{50}$, and $t_{75}$. The SFHs provide \sigmstar and \sigsfr radial profiles at both the time of observation and 1 Gyr earlier. The initial comparison shows that the shapes of the \sigsfr and \sigmstar profiles are very similar. To study star formation evolution with a quantity independent of mass or density, we therefore use projected radial profiles of log sSFR at the time of observation and 1 Gyr ago instead of \sigmstar, plotting the distribution of galaxy regions as contours over the radial profiles. 

\begin{figure*}
    \centering
    \includegraphics[width=0.9\linewidth]{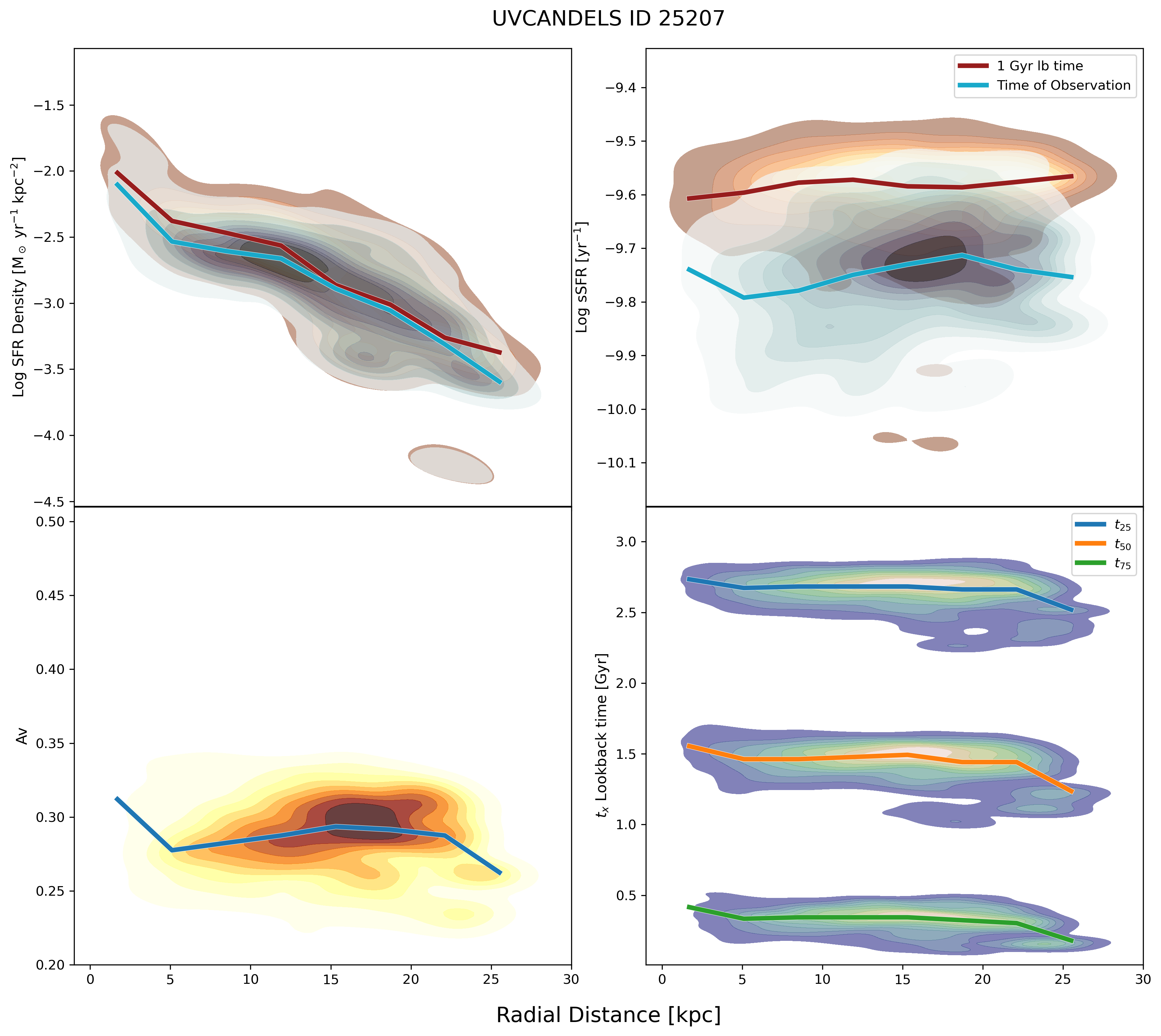}
    \caption{The projected radial profiles of galaxy 25207 have similar shapes across all properties. \sigsfr, log sSFR, Av, and $t_x$ fractions show a steep decline from the galactic center, with contours tracing the underlying distribution of regions. This behavior appears both at the time of observation and 1 Gyr lookback time, indicating a galactic center that is old yet continuously star-forming.}
    \label{fig:rad25207}
\end{figure*}

Using the redshift-dependent quenching criterion of \citep{Pacifici2016b}, we find that all regions in all galaxies lie well above the quenching threshold, indicating that both centers and outskirts are star-forming. This agrees with the strong similarity of the projected radial SFR profiles at the present time and 1 Gyr ago for most of the sample (see Fig~\ref{fig:rad25207} and Appendix~\ref{appendix:radprofiles}). Thus, for these star-forming galaxies, the SFR profiles alone do not reveal early signs of local changes in star formation.

After correcting for mass, the projected radial profiles of sSFR more clearly show the shapes and behavior of regions, as well as details from the scatter in their distribution. Precursor signs of quenching appear in most galaxies, and most of the sample shows a general sSFR decline at all radii, indicating the progressive shutdown of star formation in these massive systems. Quantitatively characterizing this decline in larger samples is a promising way to measure how rapidly star-forming massive galaxies becoming more quiescent in diverse environments, extending analyzes beyond the classic field–cluster dichotomy. This is especially timely, as large samples will soon be available from LSST, Euclid, and potential UV surveys such as GALEX \citep{Martin2005}.


\begin{figure*}
    \centering
    \includegraphics[width=0.9\linewidth]{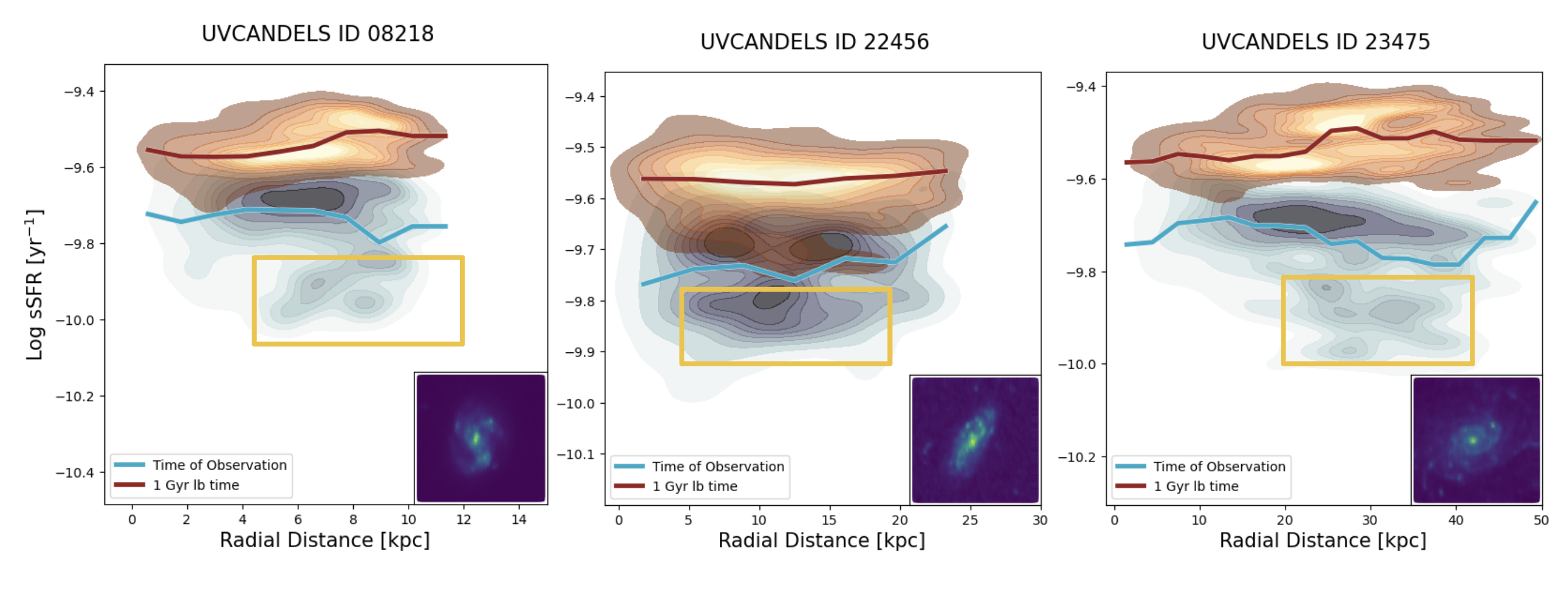}
    \caption{Projected radial sSFR profiles for UVCANDELS IDs 08218, 22456, and 23475 all show a net sSFR decline over 1 Gyr, with increasing regional scatter toward the time of observation. Boxes in each profile mark regions most rapidly trending toward quiescence..}
    \label{fig:smooth-dwindlers}
\end{figure*}

The projected radial sSFR analysis reveals increasing quiescence toward galaxy outskirts, indicating diverse environmental interactions. In Fig.~\ref{fig:smooth-dwindlers} we identify galaxies that showed slightly enhanced outer-disc star formation 1 Gyr ago, now replaced by below-average dwindling-sSFR regions at the same radii. This produces a larger spread and skewness in today’s sSFR distribution at large radii compared to 1 Gyr ago. The middle galaxy in Fig.~\ref{fig:smooth-dwindlers}, 22456, shows a more radially compact dwindling-sSFR trend, likely due to inclination; without projection effects it would resemble the more face-on 08218 and 23475. These {\it outside-in dwindlers} may be faint or early manifestations of known quenching processes, such as compressive outer-disc star formation followed by stripping-induced quenching in cluster infall \citep{Roberts2022} or fly-by stripping. This ability to detect subtler gaseous interactions between discs and their environments will, with larger samples, enable environmental studies in more diffuse intermediate-density regions—e.g. filaments and walls versus voids—predicted to strongly influence star formation \citep{Kraljic2018, Song21, Galarraga-Espinosa23, Madhani2025}.

\section{Discussion and Perspectives}
\label{sec:discussion}
We obtain informative results at $z\sim0.3$ thanks to the UVCANDELS F275W depth and angular resolution. Our galaxies are all star-forming, predominantly spirals, similar to those seen across a wide range of redshifts \citep{Kuhn2024}. 

Where previous low-redshift resolved surveys \citep[i.e.][]{Feldmann2011,ErrozFerrer2019,Pessa2021} reached sub-kiloparsec scales, we resolve structures down to four pixels, corresponding to $\approx 0.45$ kpc for the lowest-$z$ galaxies and on average $\approx 0.5$ kpc for high-SNR $z\sim0.3$ galaxies. This resolution is comparable to pixel-by-pixel SED fitting with \texttt{piXedfit} \citep{Abdurrouf2022a, Abdurrouf2022} and to the pixel-scale region sizes in \cite{Abdurrouf2025} using simulated Euclid data. 

As improved instruments and observatories push the boundary of what is considered “local,” the “semi-resolved” universe will similarly expand. Photometric SED fitting is an economical way to obtain galaxy properties, and spatially resolved SED fitting, as in this work \citep[see also][]{Cook2020, Abdurrouf2022a, Mehta2023, Abdurrouf2025}, demonstrates that photometry alone can yield meaningful results when IFU or spectroscopic data are unavailable. This approach is particularly valuable in preparation for upcoming large surveys with Euclid, Roman, and Rubin.

We recovered properties for galaxy regions consistent with observations and used reconstructed SFHs to add a temporal component to the spatially resolved SEDs, tracking each region’s evolution. We judged SFHs beyond a 1 Gyr lookback time for individual regions unreliable because of stellar mixing and assumed they revert to the galaxy’s median SFH at earlier times. We applied this cutoff to all regions, but simulations will better constrain how stellar mixing limits resolved SFH reconstruction across different areas and timescales. This is particularly important since SFHs from spatially resolved can be biased by stellar mixing and migration \citep[see][]{Minchev2025}.

\subsection{Global Trends in the Resolved SFR-M$_*$ Correlation}
\label{sec:global}

Using region SFHs, we track how properties have evolved over the past Gyr to study changes in the \sigsfr–\sigmstar correlation and infer how internal processes drive galaxy evolution.
Changes in the slope and normalization of this correlation between 1 Gyr lookback time and the time of observation can signal global quenching. Five of eight galaxies show a steeper \sigsfr–\sigmstar relation, with higher slopes but lower normalization. The three galaxies with slightly decreasing slopes (UVCANDELS IDs 10335, 22456, and 25207) host bright off-center star-forming regions, suggesting that slope flattening may be linked to rejuvenated star formation. Similar features appear in 08218 and 23475, though weaker and less prominent.

On the scale of individual regions, many begin to quench over this period, and even regions with rising SFR show declining log sSFR. Most galaxies thus become more quiescent overall, but individual regions follow diverse evolutionary paths.

Examining regional trajectories in \sigsfr and \sigmstar from 1 Gyr ago to the present reveals a wide variety of tracks, especially in galaxies with few regions where intermediate time steps (250, 500, 750 Myr) are clearly visible. In galaxies with many regions, these tracks are harder to distinguish because of crowding on the correlation. However, the global evolution in slope and normalization between 1 Gyr lookback time and today matches the aggregate regional behavior: all eight galaxies show a decrease in normalization, indicating an overall drop in \sigsfr. Many regions increase in \sigsfr over this time, but not enough to keep pace with the overall growth in \sigmstar.

\subsection{Deviations from Global Trends}
\label{sec:global}

Five of the eight galaxies show a slight increase in the slope of the \sigsfr-\sigmstar correlation between 1 Gyr lookback time and the time of observation. Two (UVCANDELS IDs 22456 and 25207) show a very slight decrease, and one (08062) remains essentially constant. 

Given the known evolution in the normalization of the global SFR–M$_*$ relation \citep[e.g.][]{Popesso2023, Speagle2014}, some evolution in its resolved counterpart is expected. Our sample spans a narrow redshift range, with the exception of UVCANDELS ID 02042 at $z=0.95$, so it is somewhat surprising that this galaxy shares both the 1 Gyr normalization and the observational-time pivot point of the resolved SFR–M$_*$ relation with the rest of the sample (Figure~\ref{fig:fit_lines}). 

The significance of the pivot at \sigmstar \~ 7.25 [$M_\odot$ kpc$^{-2}$] and \sigsfr = -2.5 [$M_\odot$ yr$^{-1}$ kpc$^{-1}$] is unclear. It may indicate a preferred stellar mass scale for star formation or reflect a region-selection metric tied to a characteristic mass. This preferred mass may indicate the presence of  self regulation tied to gravitational disk instabilities that have been shown to tightly constrain star-formation \citep{Romeo2020, Romero2020b, Romeo2023}. Reported clump masses in star-forming galaxies span 6 $\leq$ log($M_*/M_\odot$) $\leq$ 9 \citep{Guo2012, Wuyts2012, Rigby2012, Cava2018}, with lower masses from higher-resolution lensed studies. The pivot mass is broadly consistent with these clump masses and close to those in lensed systems that are less affected by outshining \citep{Harvey2025}.

 In Figure ~\ref{fig:slope_offset} we examine slope and normalization differences for individual galaxies. We first fit a line using all regions of all galaxies and centered this global fit and each galaxy’s fit at the pivot mass \sigmstar = 7.25 [$M_\odot$ $kpc^{-2}$]. We then measured the slope and normalization offsets between the global and individual-galaxy relations. Repeating this at 1 Gyr lookback time and at the time of observation, we colored points by galaxy redshift to test for redshift dependence.

\subsection{Towards hunting for early signs of quenching with SFHs?}

\label{sec:quench}

As confirmed by Fig~\ref{fig:integrated_sfrmstar}, our galaxies are star forming within a narrow redshift range with, strictly speaking, no quenched galaxies or regions. However, our sample's homogeneity allows us to show that spatially resolved SFHs can go beyond the mere search for strong past bursts or definite, irreversible quenching over the last Gyr. Indeed, while none of the galaxies in the sample has undergone a dramatic quenching event, our method was able to reveal signatures of fainter signs of past ``mini-quenching" or ``mini-bursting" phases \citep[see:][]{Dome2024, DiCesar2025} — visible as significant changes in the radial distributions of sSFR (the skewness in particular) between 1 Gyr ago and today,  coherent with detectable changes in the slopes of rSFMS correlation lines between 1 Gyr lookback time and the observation epoch.  

Quick and computationally efficient analysis of spatially resolved SFHs therefore offers a new tool to explore self-regulation and external causes of star formation in galaxies at a refined level. This is particularly timely, as new synoptic photometric surveys such as the Vera C. Rubin Observatory’s LSST and Euclid are now beginning to provide extended photometric coverage of unprecedentedly large galaxy samples across a wide variety of environments beyond the mere cluster/field dichotomy. When combined, this coverage allows for the reconstruction of spatially resolved SFHs with the fast method we presented in our study. Detailed studies of environmental interactions across large galaxy samples are therefore within reach. 

It should be noted that environmental quenching by the intra-cluster medium is already well studied in clusters \citep{Brown17, Owers19, Roberts21, Brown23}. However, less dense structures like cosmic filaments or walls are also predicted to either help preserve star formation around clusters \citep{Kotecha2022}, trigger early outside-in quenching in the field (pre-processing by large inter-cluster filaments) \citep{Aragon-Calvo19, Song21} or on the opposite boost the star-formation of low-mass galaxies \citep{Galarraga-Espinosa23} (thinner, smaller scale filaments). An issue is that observational confirmation for this complex, multi-scale interplay remains scarce due to the difficulty in detecting these fainter, at time contradictory, but prolonged  effects in action, and disentangle them from other influences. 

For instance, major integral field spectroscopy surveys show that AGN-linked quenching dominates, at the population level, in the nearby universe \citep{Penny18, Witherspoon24, Barsanti23}. However, these results in ``the field" are actually obtained from the uninformed averaging across distinct types of lower density environments with sometimes opposite effects (voids, walls, filaments bridging clusters or fainter, thinner ones). As synoptic photometric surveys now provide the statistics to study these environments separately, detecting the specific impact of mildly over-dense large-scale structures such as filaments should become easier. This will, however, require being able to observe fainter modulation of star formation over time. In this context, the outskirts of otherwise star-forming galaxies, revealed at an unprecedented level by the deep, wide-wavelength imaging from LSST and Euclid (a major advantage over current IFS surveys such as SAMI \citep{Croom12} and MaNGA \citep{Bundy2015}) are a particularly promising avenue. On a small sample, our study demonstrates that a fast, fully non-parametric SED fitting method (hence adaptable to large photometric samples) has the sensitivity to conduct such an analysis, with outskirts modulations clearly detected as significant modulations of the radial distributions of sSFR for instance, even within the context of otherwise mild quenching.

\section{Conclusion}
\label{sec:summary}
\label{sec:summary}
We performed spatially resolved analysis of galaxy evolution by mapping physical properties and measuring resolved scaling relations, enabled by the deep F275W UVCANDELS data.

Our sample includes eight GOODS-N galaxies: four face-on spirals, one edge-on, two inclined, and one pair. They are detected in all bands, have high SNR in F435W, and span spectroscopic redshifts $0.0787 \leq z \leq 0.3$. Many show bright F275W and/or F435W features indicative of recent star formation.

We constructed resolved SEDs using Voronoi binning, producing 50–800 regions per galaxy with spatial scales $\sim$0.45–7.4 kpc, set primarily by image SNR. Each region was fit with the \texttt{dense basis} SED fitting code \citep{iyer2017, Iyer2019} to obtain non-parametric SFHs and derive \sigmstar, \sigsfr, $t_{50}$, and $A_V$.

We fit the resolved $\Sigma_{SFR}$–$\Sigma_{M_*}$ relation from the regional SFHs and traced this correlation back in cosmic time. By mapping projected radial profiles of each property, we searched for temporal evolution in radial trends and patterns.

With this we found:
\begin{itemize} 
\item Using regional SFHs, we tracked each region’s evolution in \sigsfr and \sigmstar from 1 Gyr ago to the present, including 750, 500, and 250 Myr.
\item Individual regions have diverse SFHs, but galaxies show a common trend of lower normalization and overall decline in \sigsfr over the past 1 Gyr.
\item Some regions show increasing \sigsfr, but not enough to match the growth in \sigmstar, leading to an overall decrease in Log sSFR across each galaxy.
\item Comparing fits at 1 Gyr lookback time and at observation, we find strong agreement at 1 Gyr, implying all galaxies lie on the same integrated $SFR-M_*$ relation then. By the time of observation, changes in slope and normalization suggest galaxies may be quenching via different pathways. All fits pivot around \sigmstar = 7.25 [$M_\odot$ $kpc^{-2}$] and \sigsfr = -2.5 [$M_\odot$ $yr^{-1}$ $kpc^{-2}$], independent of redshift, internal scatter, inclination, or morphology, hinting at a common self-regulating scale.
\item Projected radial profiles of galaxy properties have similar shapes across $t_{x}$ fractions, indicating little stellar radial migration.
\item Comparing 1 Gyr lookback time and present-day radial profiles of \sigsfr and Log sSFR shows that most galaxies undergo an overall decline in Log sSFR, with larger present-day sSFR scatter indicating some regions are moving faster toward quiescence.
\end{itemize}

Our analysis of the correlation lines for each galaxy indicates a preferential \sigmstar of 7.25 [$M_\odot$ $kpc^{-2}$] at the time of observation and that all eight galaxies shared the same integrated $SFR-M_*$ correlation at 1 Gyr lookback time. This \sigmstar value suggests that further study could clarify which physical processes dominate at small scales in MW-type galaxies. 
Projected radial profiles of galaxy properties at different time-steps, enabled by spatially resolved SFHs, reveal the spatial evolution of galaxy regions and highlight where quenching may begin, even in star-forming systems. Applying this analysis to large surveys such as LSST, Euclid, and Roman will deepen our understanding of how and when star formation regulates galaxy evolution. 

We demonstrated that our data and code can recover resolved properties and test resolved scaling relations, but this work only begins to explore the science accessible with SFH reconstruction on resolved datasets. We reliably recovered regional properties from SEDs with our nonparametric SFH reconstruction code, tracked the evolution of \sigsfr and \sigmstar, and observed changes in slopes and normalizations over time. This approach opens rich avenues for future work, including analysis of projected radial trends, coevolution of adjacent regions, resolved dust and metallicity, and constraints on stellar mixing. 
Our method is consistent with existing spatially resolved techniques such as IFU, spectroscopic, and pixel SED analyses, and can quickly and efficiently be applied to diverse datasets while adding temporal information through SFH reconstruction. Applying these techniques to upcoming surveys will enable a deeper exploration of the evolution of spatially resolved regions in galaxies and yield sharper insights into the location and timing of internal processes that drive galaxy growth.

\section{Acknowledgments}
We thank the anonymous referee for valuable suggestions that improved this work. The analysis presented in this paper is based on observations with the NASA/ESA Hubble Space Telescope obtained at the Space Telescope Science Institute, which is operated by the Association of Universities for Research in Astronomy, Incorporated, under NASA contract NAS5-26555. Support for Program Number HST-GO-15647 was provided through a grant from the STScI under NASA contract NAS5-26555.  CO and EG acknowledge support from NASA Astrophysics Data Analysis Program grant 80NSSC22K0487. CO and CW are supported by an LSST-DA Catalyst Fellowship funded through Grant 62192 from the John Templeton Foundation to LSST Discovery Alliance and CW is supported by The NSF LEAPS-MPS award 2316862. XW is supported by the National Natural Science Foundation of China (grant 12373009), the CAS Project for Young Scientists in Basic Research Grant No. YSBR-062, the Fundamental Research Funds for the Central Universities, the Xiaomi Young Talents Program, and and the China Manned Space Program with grant no. CMS-CSST-2025-A06. RAW acknowledges support from NASA JWST Interdisciplinary Scientist grants NAG5-12460, NNX14AN10G and 80NSSC18K0200 from GSFC. Y.S.D acknowledges the support from the National Natural Science Foundation of China grant No. 12273051.

\bibliography{thesis}{}
\bibliographystyle{aasjournalv7}

\appendix
\vspace{-5mm}
\section{Property maps}
\label{appendix:propmaps}

\begin{figure}[H]
    \centering
    \includegraphics[width=0.9\linewidth]{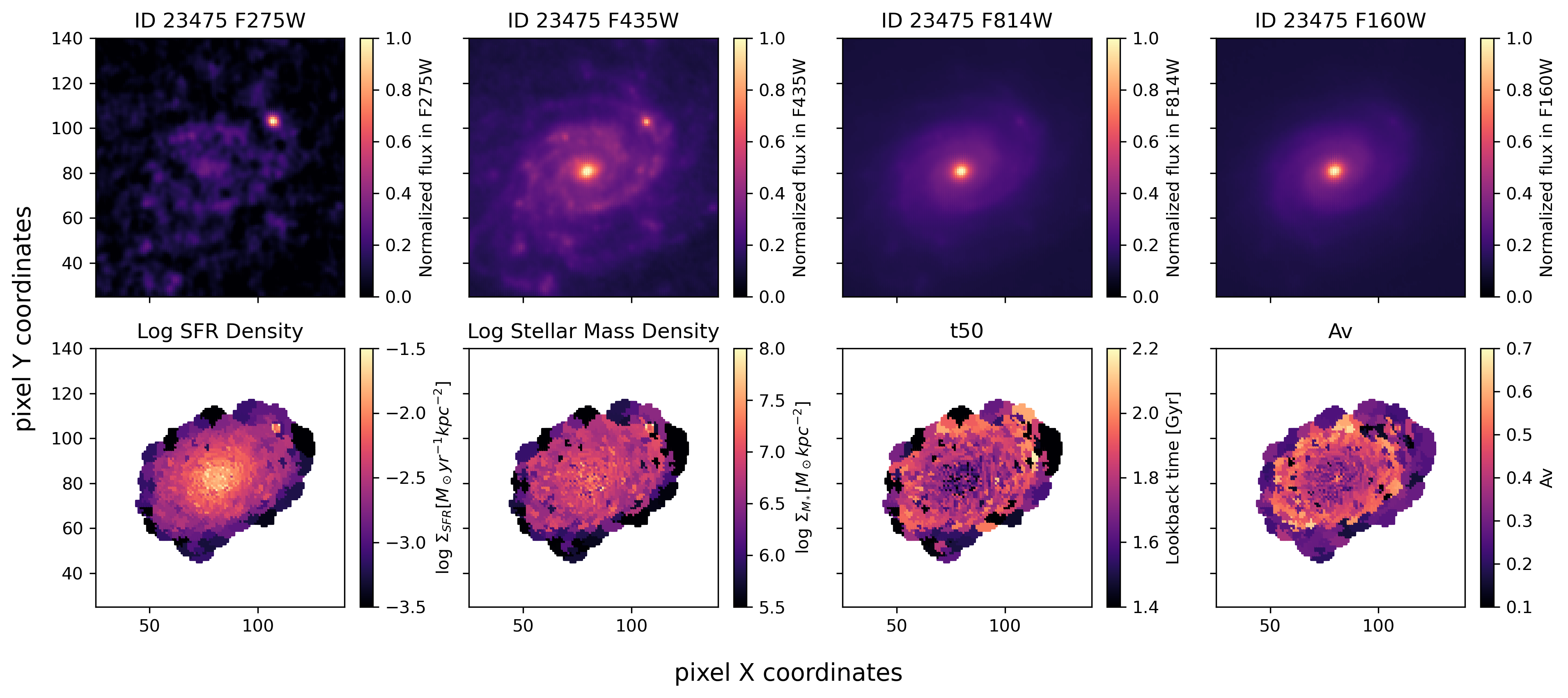}
    \caption{Within the UVCANDELS 23475 postage stamps, the galaxy’s spiral arms are clearest in F435W. The bright star-forming knot in the upper right is prominent in SFR density and faint in stellar mass density. The $t_{50}$ map shows this region has a similar age to the rest of the galaxy. The $A_V$ map is fairly smooth, with less dust near the bulge than in the disc.}
    \label{fig:multi_23475}
\end{figure}

We present the property maps for the remaining six galaxies in our sample. Figures~\ref{fig:multi_02042} to ~\ref{fig:multi_23475} use the same format as the three example galaxies, showing F275W, F435W, F814W, and F160W images above maps of \sigsfr, \sigmstar, $t_{50}$, and $A_V$.

\begin{figure}[H]
    \centering
    \includegraphics[width=0.9\linewidth]{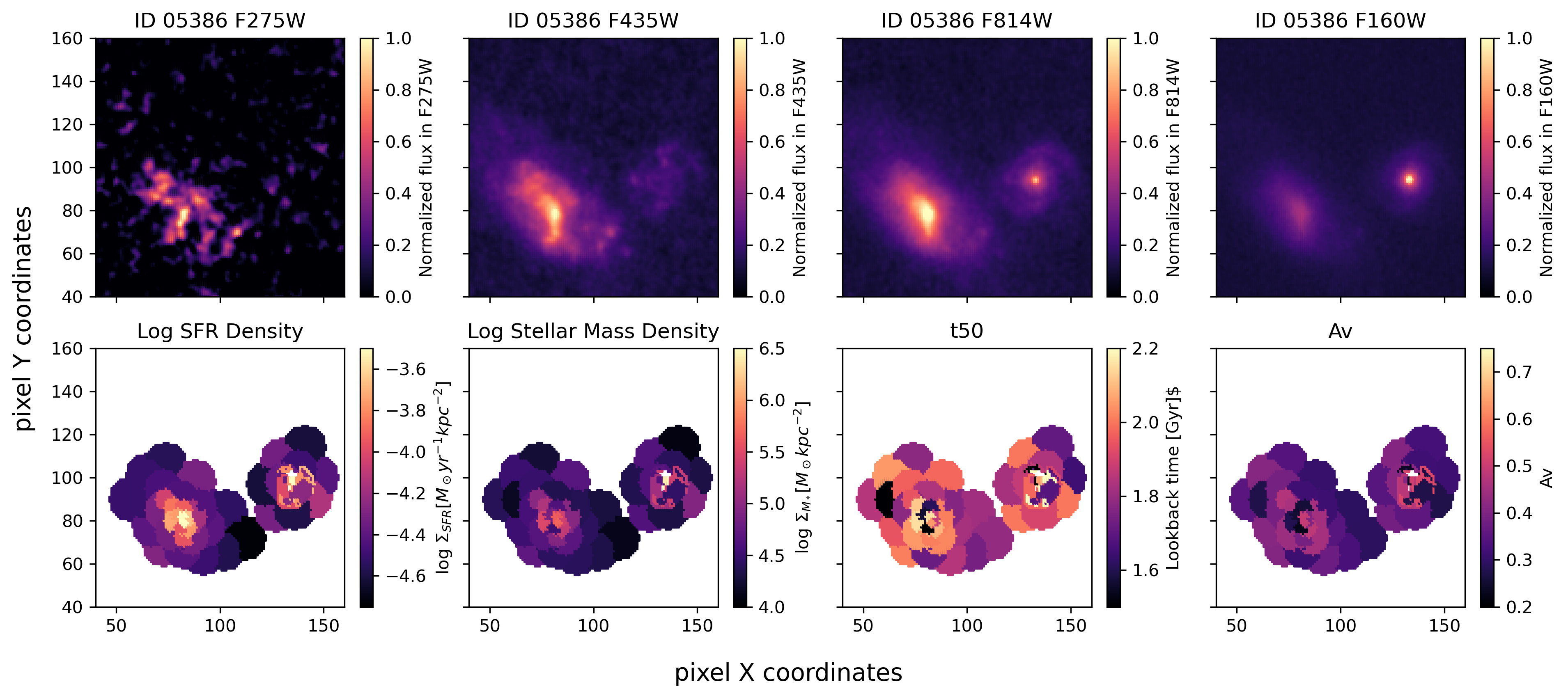}
    \caption{UVCANDELS 05386 is cataloged as a pair, with the dimmer object to the right likely a companion. In F814W and F160W, two distinct objects are evident, although only the larger lower-left galaxy is bright in F275W. Both objects appear in the F435W segmentation map. SFR density peaks at the centers of both galaxies, while stellar mass is generally low except in these centers. The $t_{50}$ map indicates younger regions in both the outskirts and centers, and $A_V$ is relatively uniform.}
    \label{fig:multi_05386}
\end{figure}

Figure ~\ref{fig:multi_05386} shows postage stamps for UVCANDELS ID 05386 with 50 regions defined before the SNR cutoff. The images contain two galaxies at z = 0.1363, confirmed as a pair by \cite{ORyan2023ApJ}. The lower-right galaxy is bright in F275W and F435W, while the other is brighter in redder bands, suggesting it is either more quiescent or at a different redshift than its apparent companion.

In the segmentation maps, the SFR in the lower galaxy is less extended than the F275W emission suggests. The galaxy to the right shows elevated SFR despite low F275W flux. The stellar mass density map for 05386 is mostly uniform, with low masses that rise toward the center.
\vspace{-5mm}
\begin{figure}[H]
    \centering
    \includegraphics[width=0.9\linewidth]{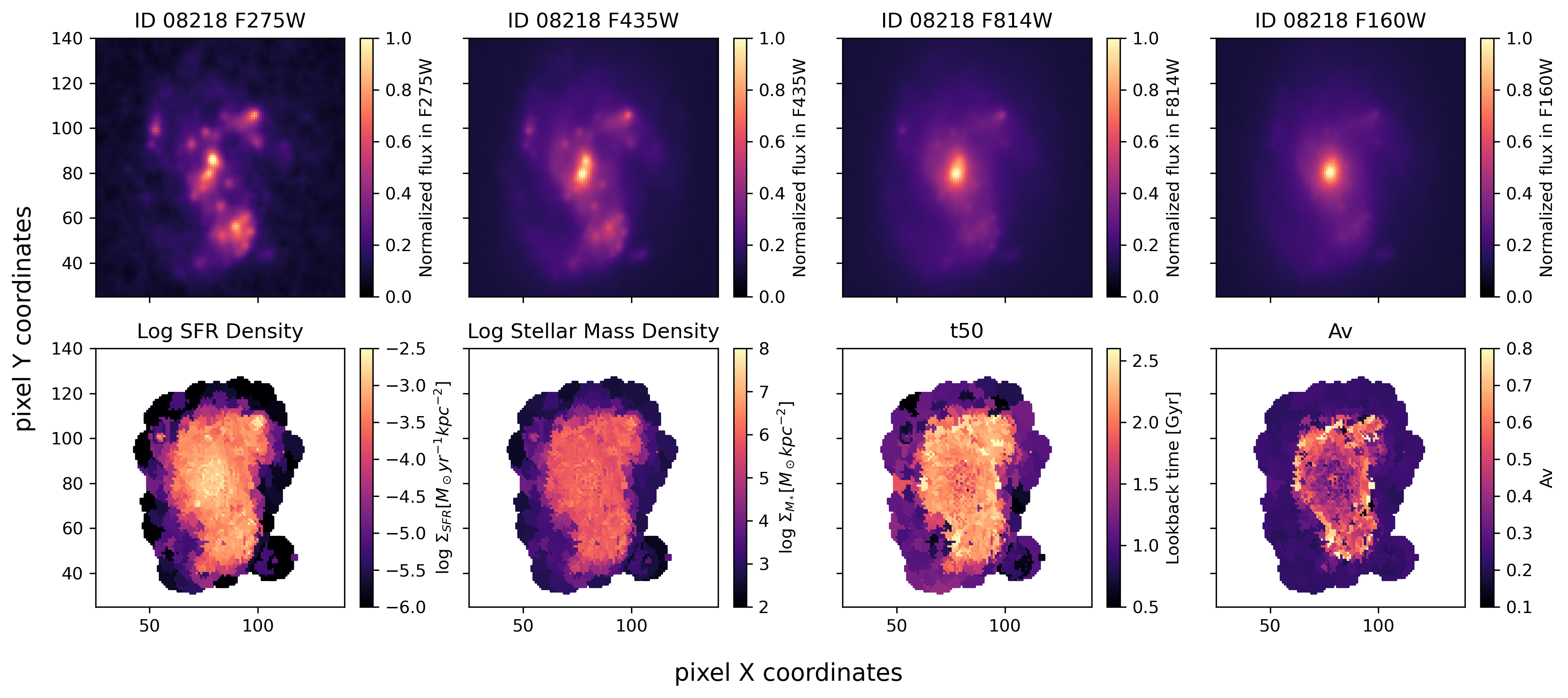}
    \caption{Shown is UVCANDELS 08218. Clear star-forming clumps appear in F275W and remain visible in F435W, but are hard to identify in the segmentation map. In the SFR density map, some features are visible on close inspection, with elevated SFR density in the upper right and upper left regions of the galaxy. The stellar mass density and $t_{50}$ show little variation, while much of the galaxy has high $A_V$.}
    \label{fig:multi_08218}
\end{figure}

\vspace{-10mm}

Figure ~\ref{fig:multi_08218} shows UVCANDELS ID 08218, a face-on spiral with clear clumps in F275W and F435W. These clumps are not evident in the property maps except for \sigsfr, despite the 750 defined regions. Although more regions do not necessarily yield more detail, the \sigsfr map shows enhanced values in the upper right and upper left of the galaxy. Overall, the image structure is weakly reflected in the segmentation maps; dust and $t_{50}$ in particular show little correspondence with the postage stamps.


\begin{figure}[H]
    \centering
    \includegraphics[width=0.9\linewidth]{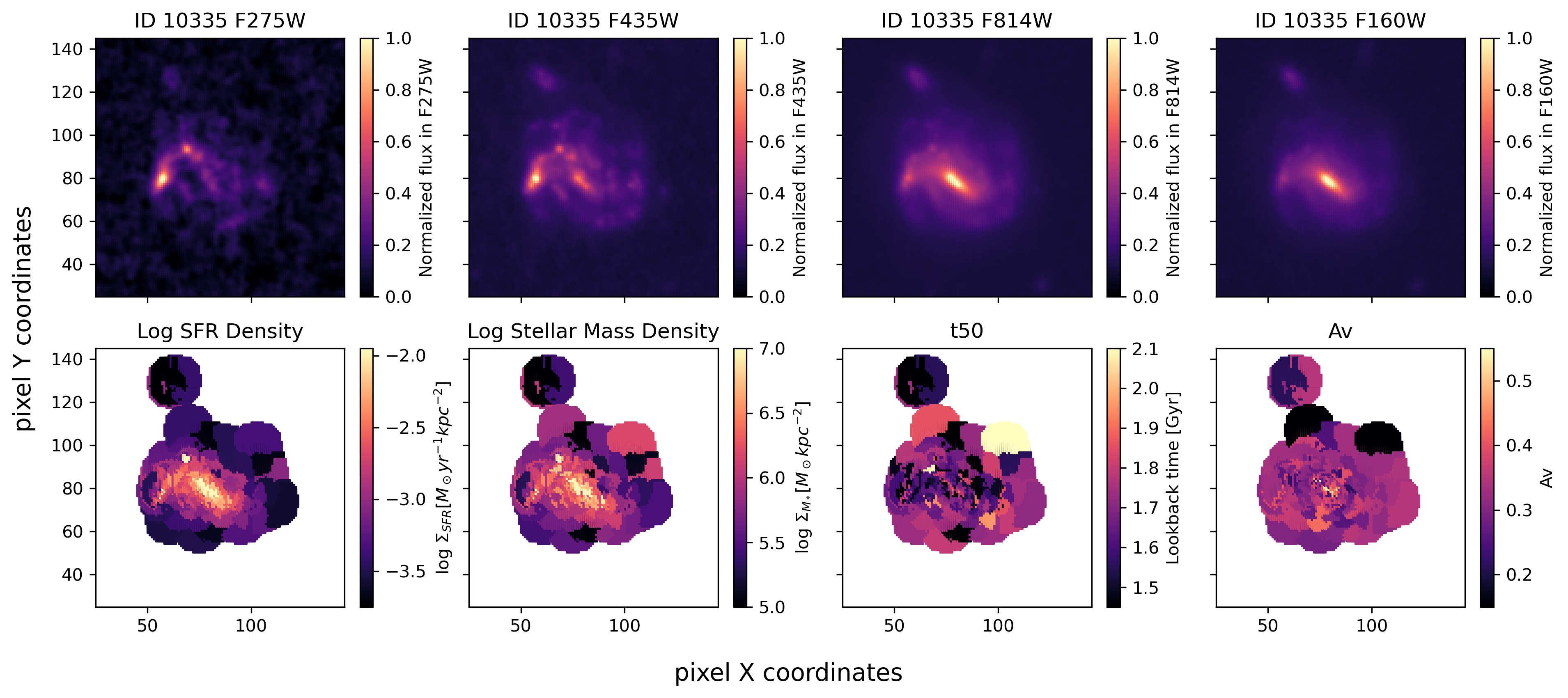}
    \caption{Shown are postage stamps and property maps of 10335 in pixel coordinates. \textbf{Top}: HST F275W, F435W, F814W, and F160W postage stamps. \textbf{Bottom}: 2D maps of log $\Sigma_{SFR}$, $\Sigma_{M_*}$, $t_{50}$, and $A_V$. The warped or twisted feature is clear in the segmentation maps and especially in $\Sigma_{SFR}$ and $\Sigma_{M_*}$, but not in $t_{50}$ or $A_V$. A bright knot at the top of the twisted arm in F275W and F435W stands out in the $\Sigma_{SFR}$, $\Sigma_{M_*}$, and $t_{50}$ maps, indicating an older, more massive region currently undergoing elevated star formation. The object in the upper left corner is very faint in all properties, suggesting it is a background source unrelated to 10335.}
    \label{fig:multi_10335}
\end{figure}
UVCANDELS ID 10335 is a spiral galaxy with a twisted arm. Before applying our SNR cutoff, the algorithm identified 160 regions that capture the galaxy’s visible structure, including the faint feature in the upper left of Figure ~\ref{fig:multi_10335}. Although faint in all bands, this feature appears in the segmentation and may be a clump of tidally disturbed gas, an older object, or a background source. The galaxy’s main structures—especially the bright edge of the twisted arm in the lower left—are prominent in the property maps. They are most clearly traced in \sigsfr and \sigmstar, while $t_{50}$ is more irregular and $A_V$ shows little variation. A particularly notable feature is a bright knot above and slightly left of the galactic center, clearly visible in the F275W band. It lies on or just above the twisted arm, is especially bright in \sigsfr and \sigmstar, and may indicate tidally triggered star formation due to an interaction or merger.

\begin{figure}[H]
    \centering
    \includegraphics[width=0.9\linewidth]{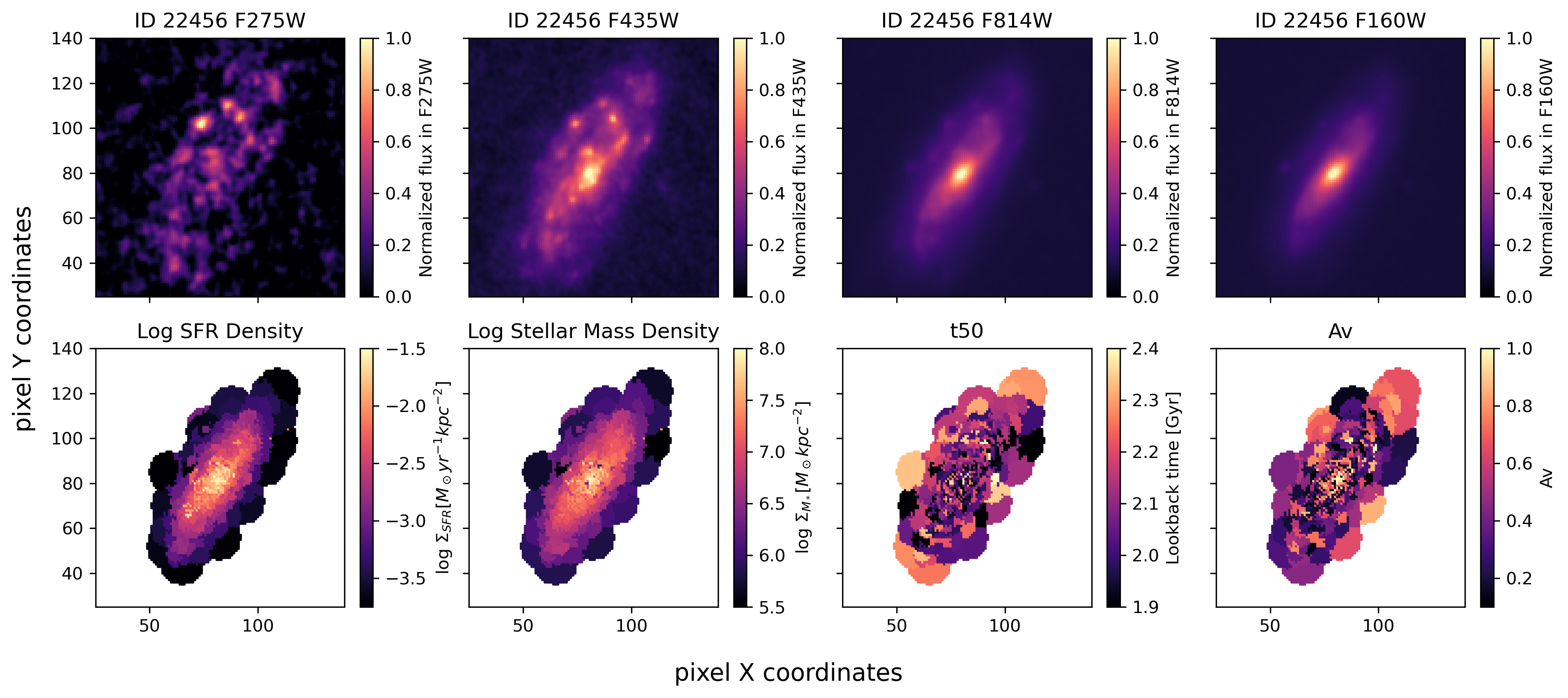}
    \caption{UVCANDELS 22456 is an inclined spiral galaxy. Strong clumps trace the spiral arms in F435W and are also clear in F275W. These clumps are not evident in SFR or stellar mass density maps, though the brightest F275W clump appears only faintly in SFR density. The $t_{50}$ and $A_V$ maps show no clear patterns, but $A_V$ is higher in the brightest regions, which may explain the weak SFR inferred from SED fits there. }
    \label{fig:multi_22456}
\end{figure}

UVCANDELS ID 22456 is shown in Figure~\ref{fig:multi_22456}. In the top row, the spiral arms are visible in all bands, with bright clumps in F435W and F275W. The tesselation produced 375 segments, colored by property in the second row. None of the property maps reproduce the spiral arms, though smaller features appear in \sigsfr and \sigmstar. The bright knots in F275W are visible in both \sigsfr and \sigmstar, faintly in the galaxy’s upper left. Because we do not “de-zone” the image (e.g., luminosity-weight pixels within a region), it is not visually obvious in the \sigsfr and \sigmstar maps that these clumps are captured by the regions, as they lie in zones that also contain lower \sigsfr and \sigmstar pixels. 

The $t_{50}$ map follows the UV (F275W) structure: regions around bright F275W knots and those covering dark F275W areas are youngest. Bright knots only faintly seen in \sigsfr and \sigmstar appear relatively young, consistent with the UV band tracing young star-forming clumps. The $A_V$ map roughly matches the galaxy’s features, clearly showing the brightest knot and identifying it as one of the dustiest regions. If that were true, we would still clearly see this bright feature in \sigsfr given its UV flux. Thus, $A_V$ may be poorly constrained in this galaxy, likely causing \sigsfr to be underestimated.

\vspace{-10mm}

UVCANDELS ID 23475 is shown in Figure~\ref{fig:multi_23475}. The postage stamps show a relatively smooth spiral galaxy in F814W and F160W. F435W and F275W show a bright knot along a spiral arm in the upper right quadrant and flocculant features along the arms. The property maps show a more uniform distribution, with \sigsfr forming a smooth gradient from the center. The bright knot stands out in \sigsfr and is also visible, though less prominently, in \sigmstar. The $t_{50}$ map indicates that the center formed 50\% of its mass earlier than the outskirts. The bright knot has an average age and is bordered by very young and very old regions. The $A_V$ map shows lower dust in the outskirts and near the center, and is otherwise uniform.

\section{Evolving Correlations}
\label{appendix:sfrmstar}
\begin{figure}[H]
    \centering
    \includegraphics[width=0.9\linewidth]{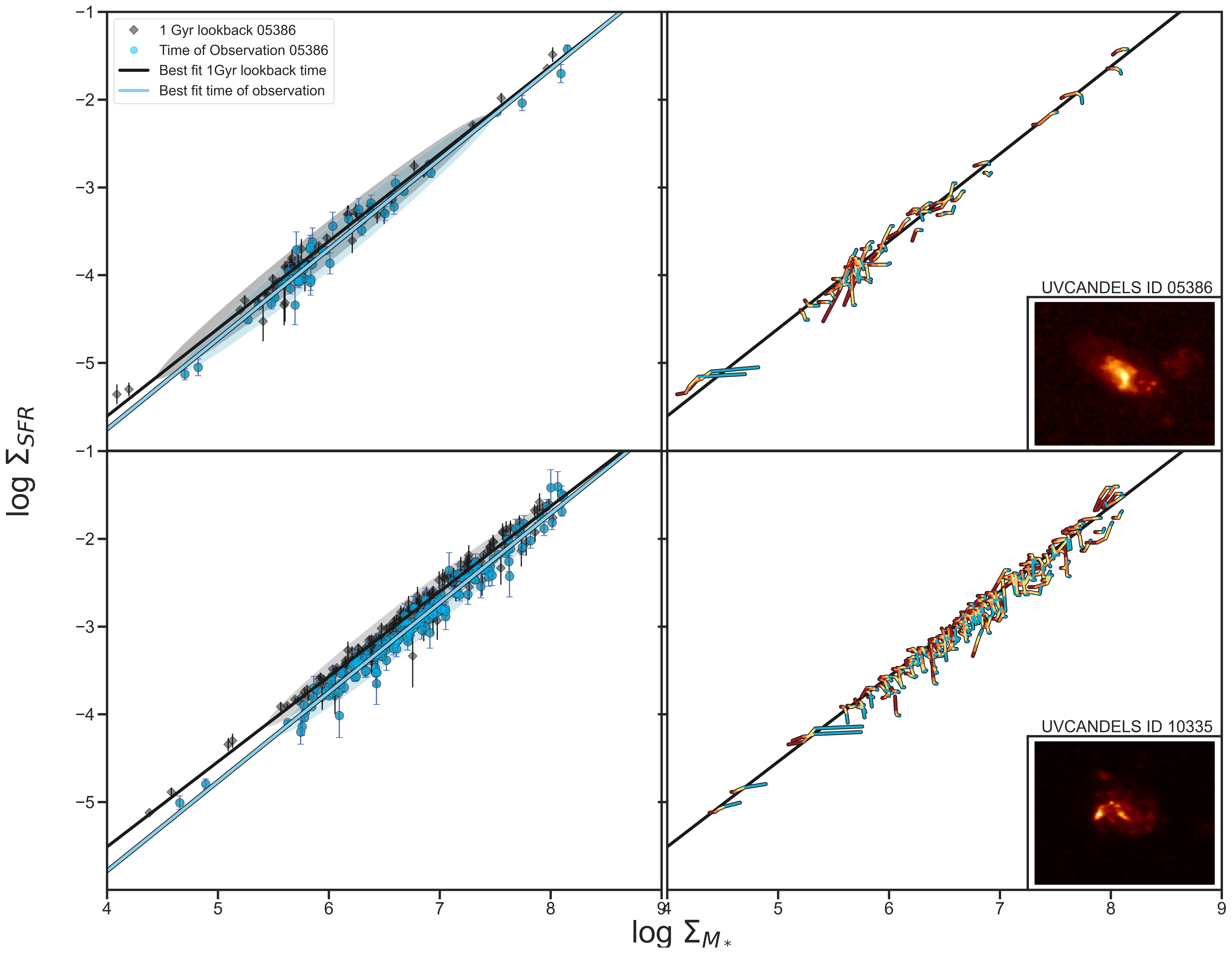}
    \caption{Evolving \sigsfr–\sigmstar correlations for UVCANDELS IDs 05386, and 10335. Left: points and trend lines at the time of observation (cyan) and 1 Gyr lookback time (black). Right: more detailed evolution with additional time steps at 750, 500, and 250 Myr. These galaxies have relatively few regions, making their diverse evolutionary tracks clearly visible. All three show minor changes in slope and normalization in their correlation lines..}
    \label{fig:sfr_mstar_lownum}
\end{figure}

\begin{figure}
    \centering
    \includegraphics[width=0.9\linewidth]{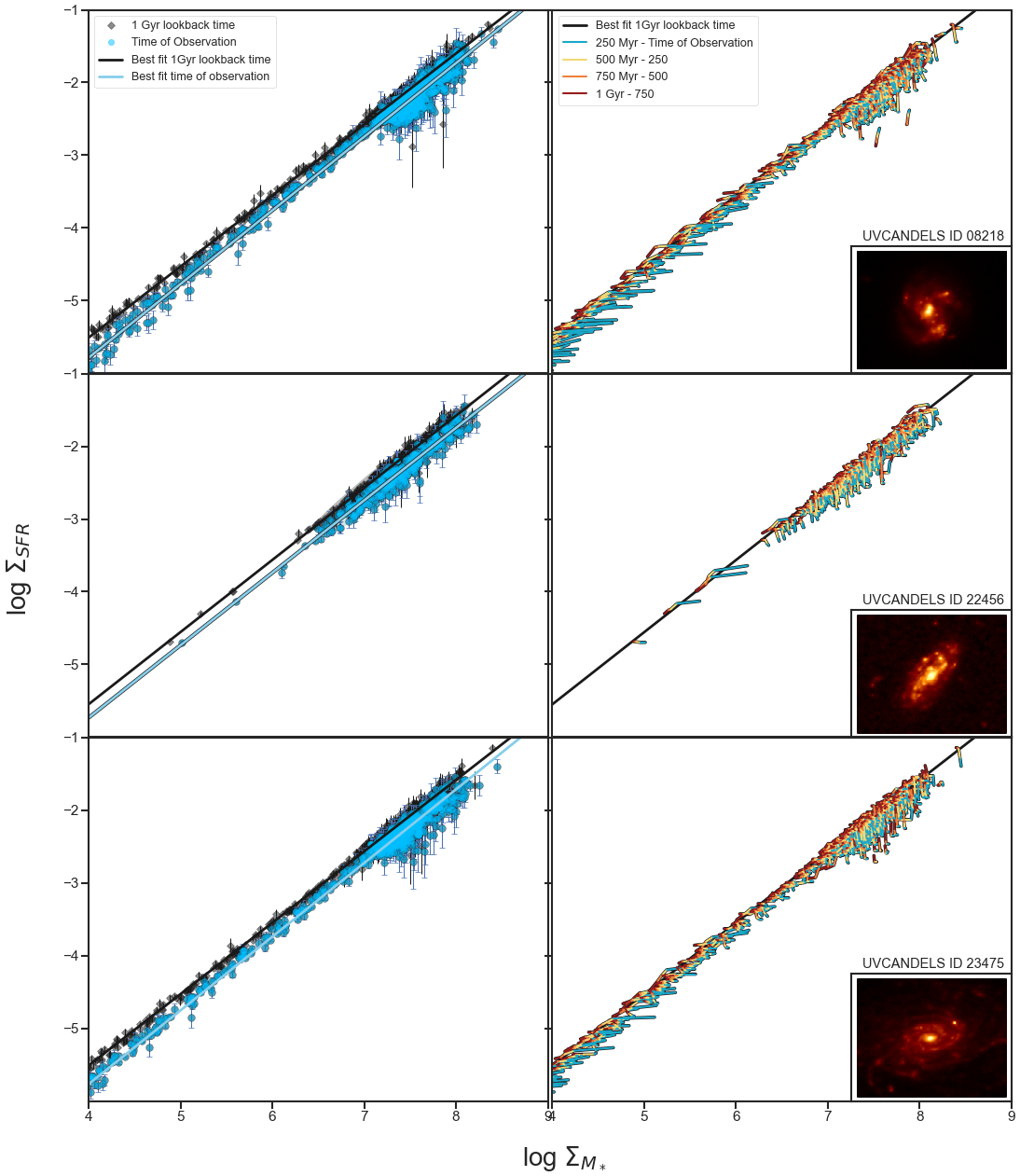}
    \caption{Similar to Figure~\ref{fig:sfr_mstar_lownum}, we show the evolution of regions within UVCANDELS IDs 08218, 22456, and 23475. These galaxies contain more regions, and their high density obscures the variety of evolutionary tracks seen in other galaxies. UVCANDELS 08218 and 23475 in particular show many regions with \sigmstar $\sim$7–8.5 [$M_\odot$ $kpc^{-2}$], whose \sigsfr typically declines steeply from a 1 Gyr lookback time time.}
    \label{fig:sfr_mstar_highnum}
    \vspace{2cm}
\end{figure}

Figures~\ref{fig:sfr_mstar_lownum} and ~\ref{fig:sfr_mstar_highnum} show the \sigsfr \sigmstar correlations and evolution of regions over 1 Gyr for the remaining galaxies. We see in Figure~\ref{fig:sfr_mstar_highnum} galaxies with as many as 800 regions, while the galaxies in Figure~\ref{fig:sfr_mstar_lownum} have between 49 and 160 regions. 

Figure~\ref{fig:sfr_mstar_lownum} shows that although each galaxy’s slope and normalization decrease from 1 Gyr lookback time to the time of observation, regional evolutionary tracks follow diverse SFHs in \sigsfr-\sigmstar space. Galaxies with more regions (Figure~\ref{fig:sfr_mstar_highnum}) exhibit stronger trends in these tracks. Most regions clearly decline in \sigsfr over the Gyr as \sigmstar increases, but not enough to remain on the 1 Gyr lookback time correlation. 
\clearpage

\vspace{4cm}
\section{Radial Profiles}
\label{appendix:radprofiles}
\begin{figure}[h]
    \centering
    \includegraphics[width=0.6\linewidth]{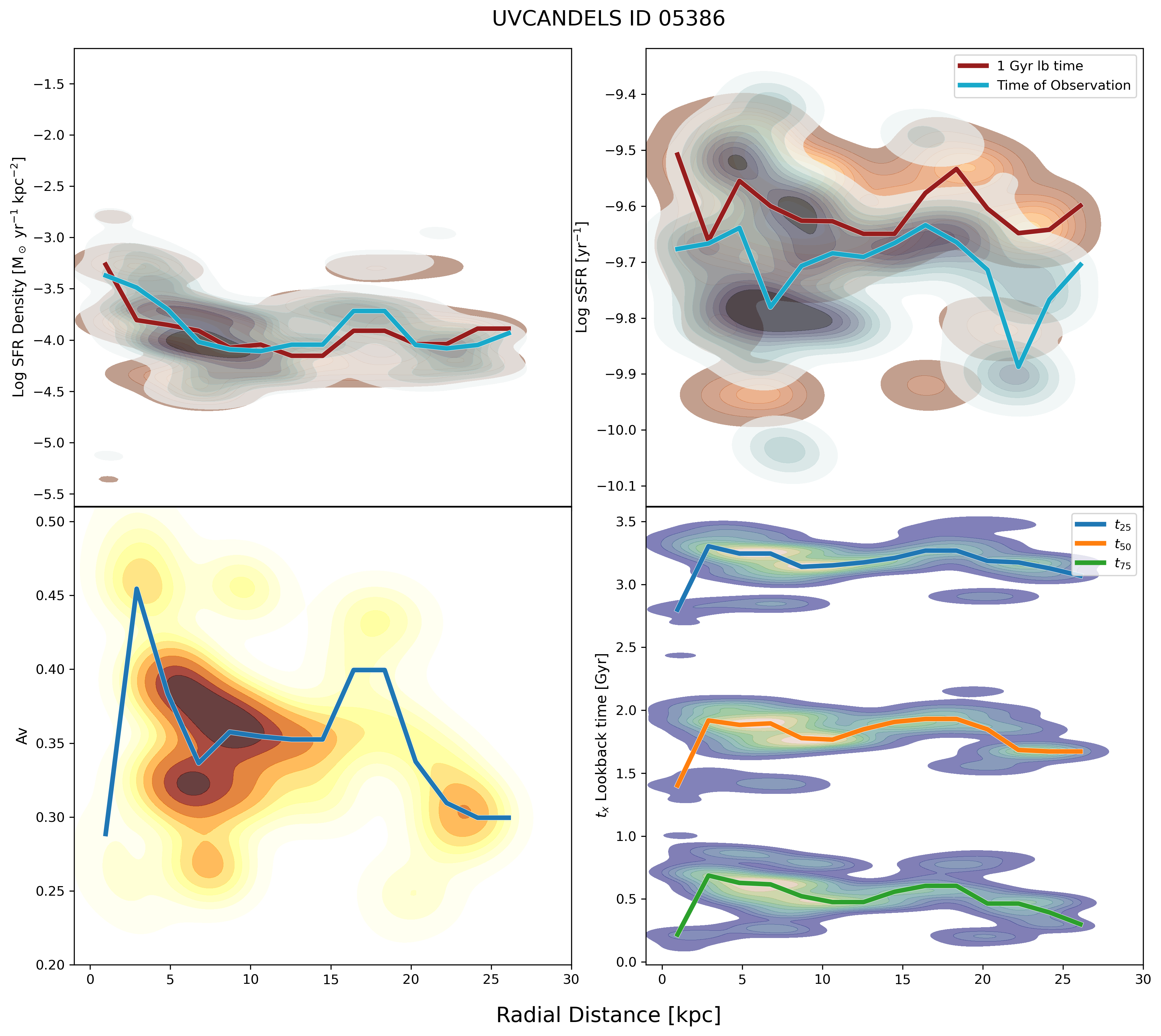} 
    \caption{The “center” of UVCANDELS ID 05386 is defined as the brightest point in the object on the left. Because this is actually two distinct objects, the projected radial profiles appear somewhat scrambled, but the \sigsfr and log sSFR still agree over the last 1 Gyr.}
    \label{fig:rad05386}
\end{figure}

\begin{figure}
\vspace{0.2 cm}
\begin{minipage}[t]{0.49\linewidth}\vspace{0pt}
\begin{flushleft}
     \includegraphics[width=\linewidth]{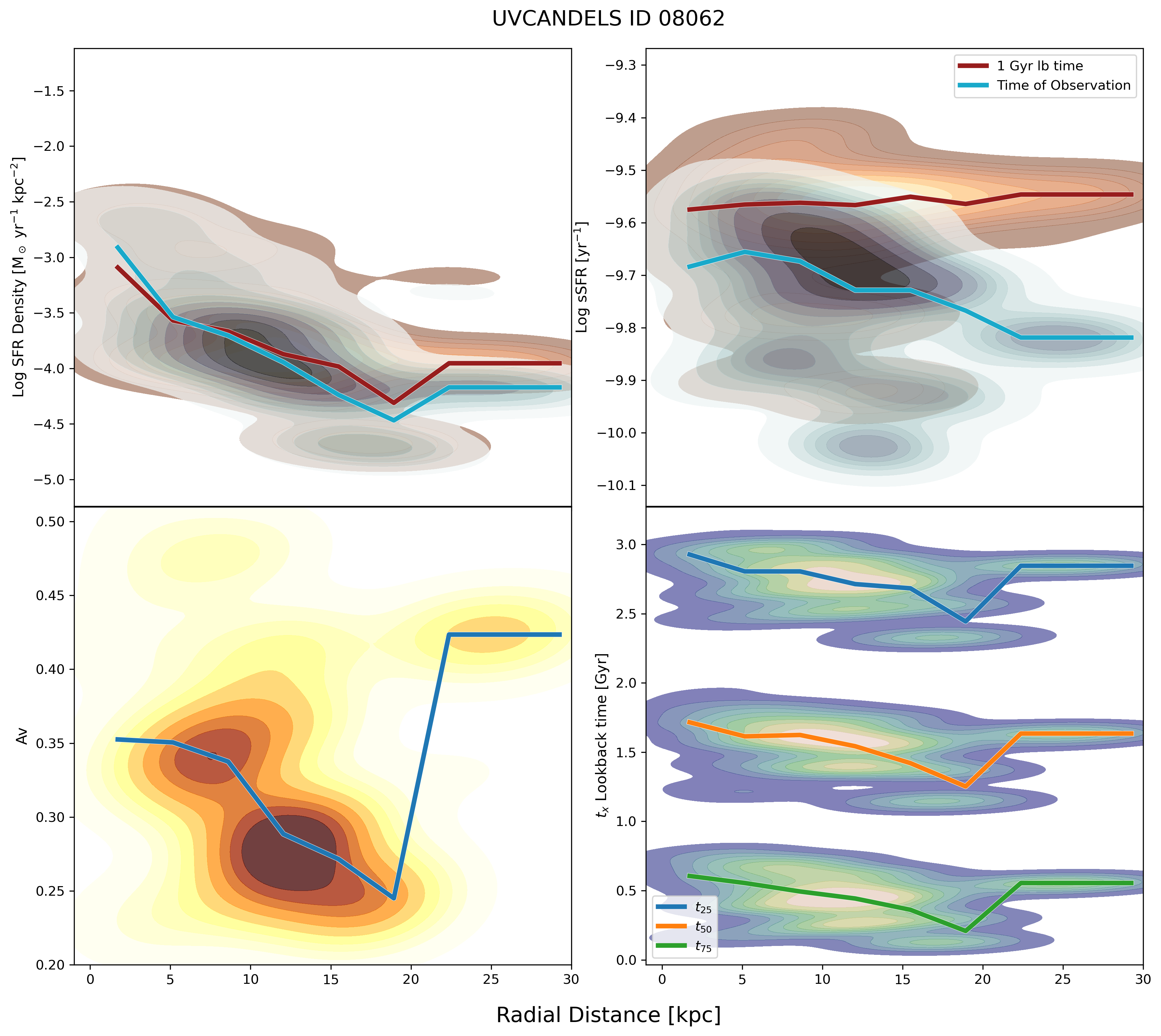}
    \caption{Radial profiles for 08062 in \sigsfr, log sSFR, $A_V$, and $t_x$ mass formation times show similar \sigsfr\ shapes at both time steps, indicating little radial migration. Notably, the observed log sSFR is lower than that 1 Gyr ago in the outskirts, providing early evidence for outside-in quenching.}
    \label{fig:rad08062}
\end{flushleft}
\end{minipage}
\begin{minipage}[t]{0.49\linewidth}\vspace{0pt}
\begin{flushright}
        \includegraphics[width=\linewidth]{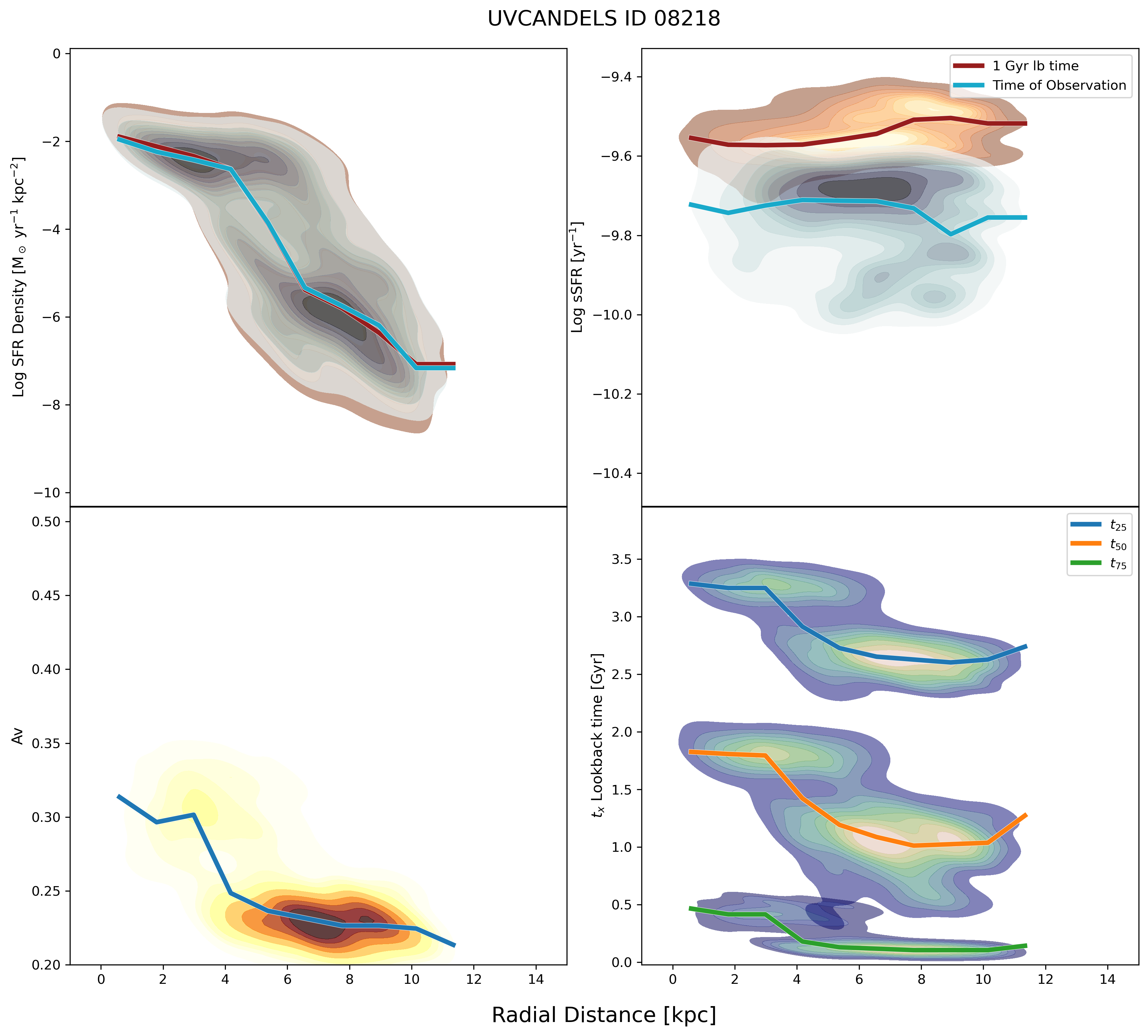}
        \caption{The \sigsfr radial profile for UVCANDELS ID 08218 shows little evolution over the last Gyr in any region, but the $\log$ sSFR profile indicates a slight decline in star formation across the entire galaxy.}
    \label{fig:rad08218}
\end{flushright}
        
\end{minipage}

\end{figure}

\begin{figure}
\vspace{0.2 cm}
\begin{minipage}[h]{0.49\linewidth}
\begin{center}
\includegraphics[width=\linewidth]{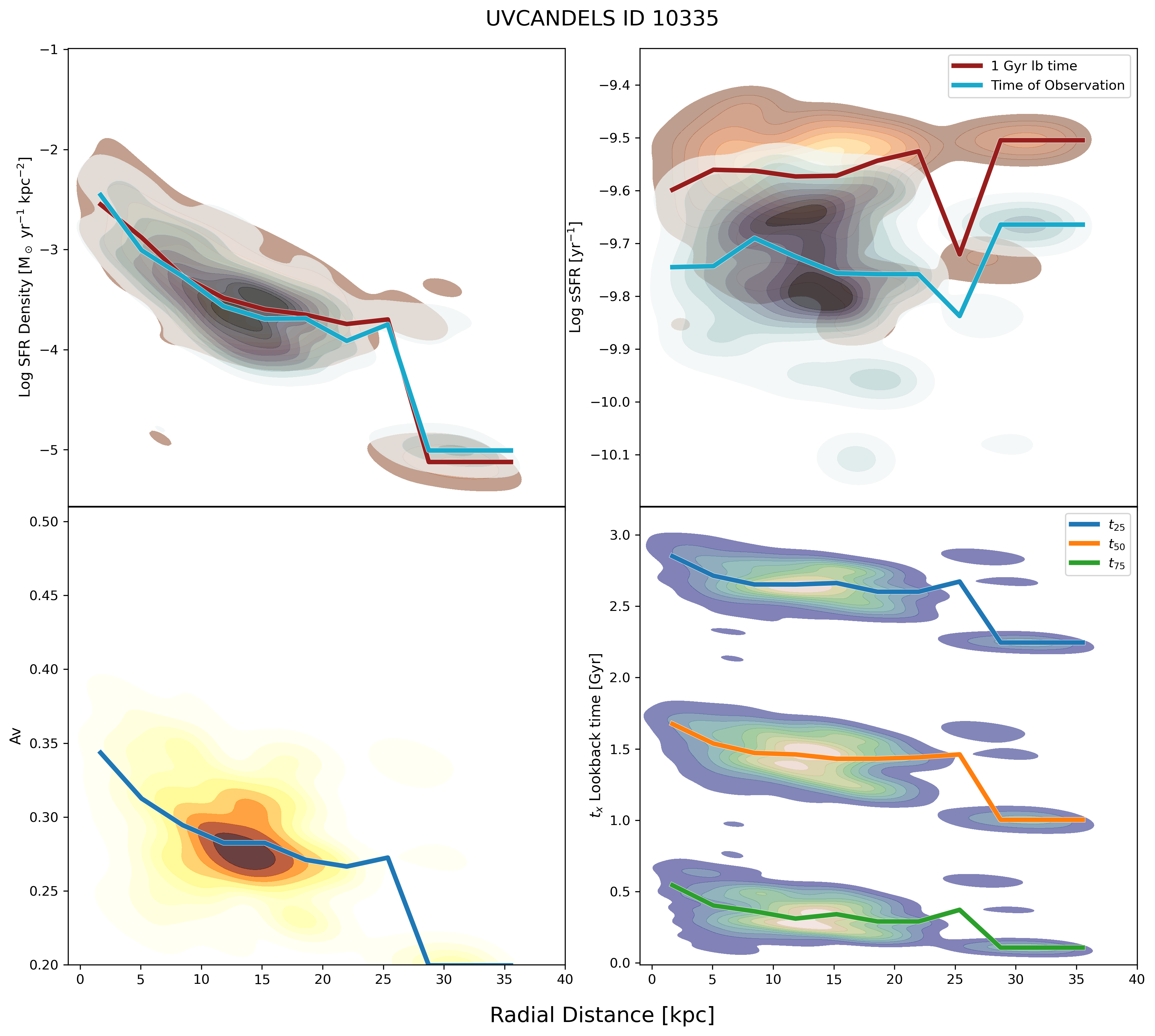}
    \caption{The projected radial profile of UVCANDELS ID 10335 shows similar shapes for all properties: \sigmstar, \sigsfr, Av, and $t_x$ are highest at the galactic center and decline out to 20 kpc. Around 22 kpc, all properties rise again before dropping sharply in the outskirts. This increase corresponds to the bright star-forming knot in the postage stamps, and the outer depression aligns with the tidal feature beyond the disk. }
    \label{fig:rad10335}
    
\end{center}
\end{minipage}
\hfill
\begin{minipage}[h]{0.49\linewidth}
\begin{center}
\includegraphics[width=\linewidth]{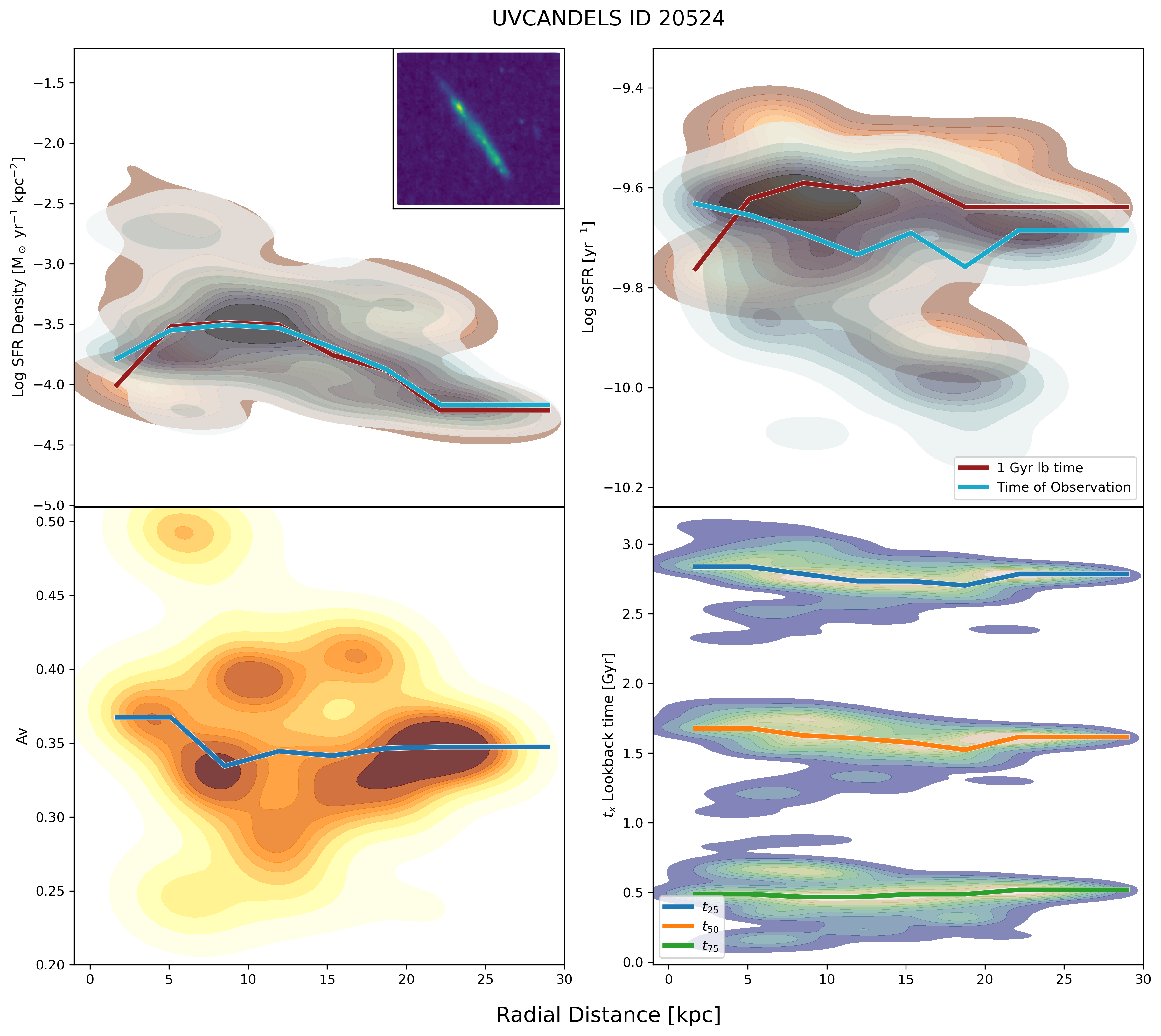}
   \caption{Both \sigsfr and log sSFR show a crossing behavior where the values at the time of observation exceed those at 1 Gyr lookback time within the inner 5 kpc. This likely results from projection effects in the edge-on galaxy. \\ \\ \\ \\ }
   \label{fig:rad20524}
\end{center}
\end{minipage}
\end{figure}

\begin{figure}
\vspace{0.2 cm}
\begin{minipage}[h]{0.49\linewidth}
\begin{center}
\includegraphics[width=\linewidth]{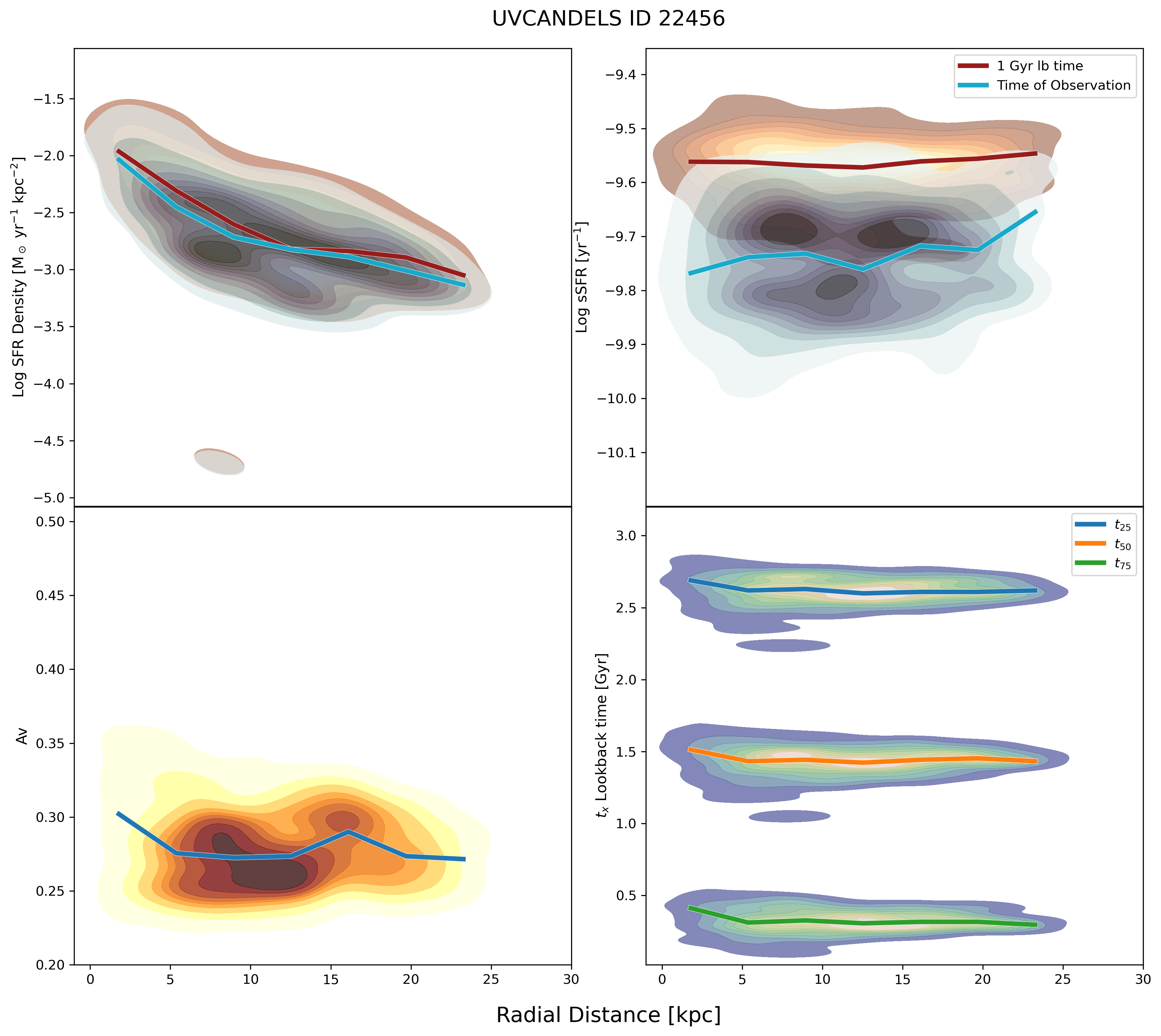}
    \caption{The projected radial profiles of UVCANDELS ID 22456 show a fairly uniform decrease across the galaxy in \sigsfr and Log sSFR. Regions in sSFR show a diffuse spreading and overall drop at the time of observation suggesting possible early signs the galaxy is trending towards quenching in these regions. }
    \label{fig:rad22456}
\end{center}
\end{minipage}
\hfill
\begin{minipage}[h]{0.49\linewidth}
\begin{center}
    \includegraphics[width=\linewidth]{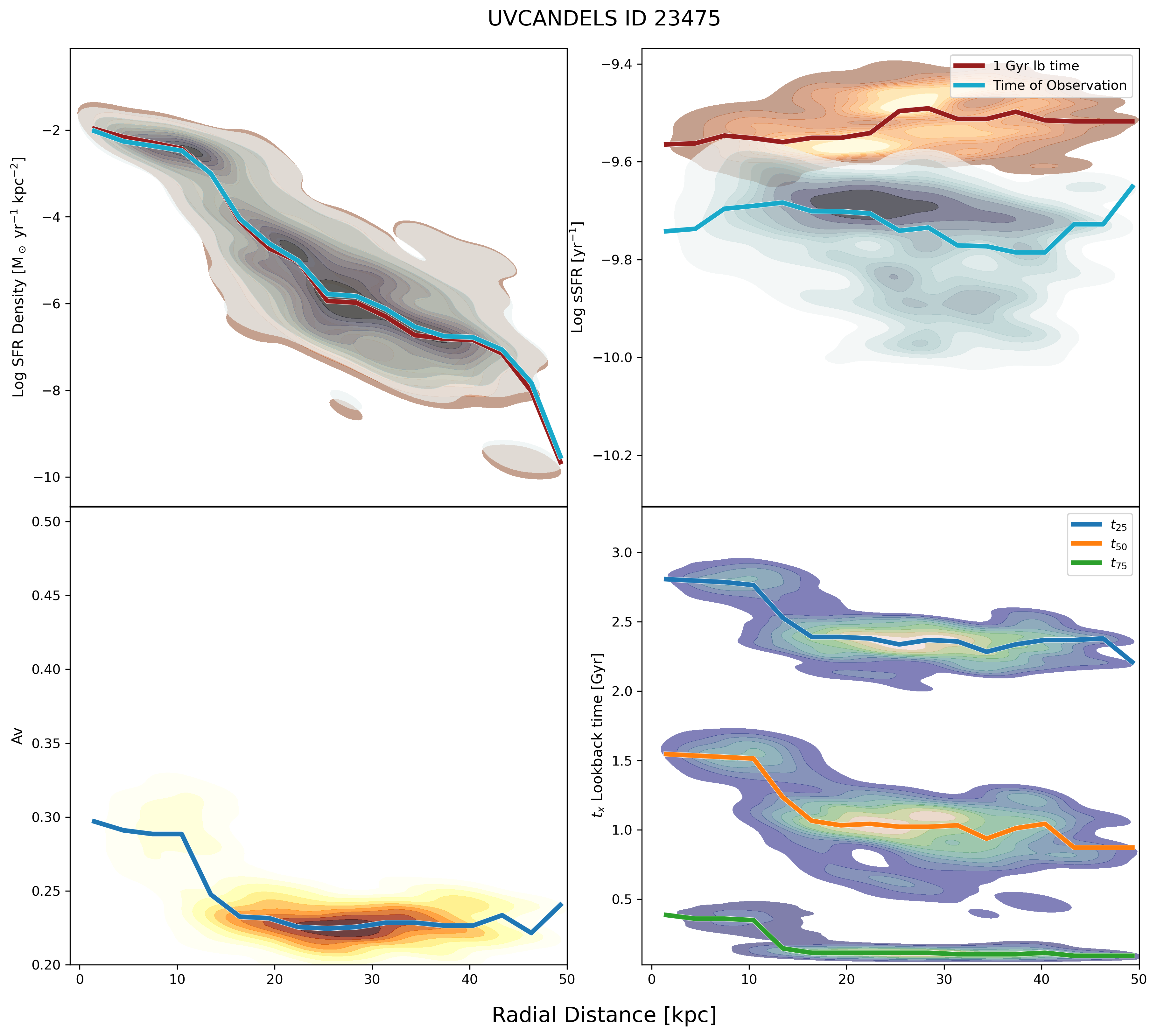}
    \caption{The projected radial profiles for log sSFR show that all regions have lower star formation at the time of observation, with increased scatter and many regions falling to even lower values suggesting regions may quench in the future. \\ \\ }
    \label{fig:rad23475}
\end{center}
\end{minipage}
\label{ris}
\end{figure}

Figures~\ref{fig:rad05386}–\ref{fig:rad23475} show the projected radial profiles of the remaining galaxies. Each figure presents the projected radial profiles of \sigsfr, Log sSFR, $A_V$, and the $t_x$ fractions $t_{25}$, $t_{50}$, and $t_{75}$. Overall, the $t_x$ profiles have similar shapes, implying limited radial migration, and \sigsfr is elevated near the galactic center relative to the outskirts. Because the \sigsfr and \sigmstar profiles closely match in shape and behavior, we show Log sSFR instead of \sigmstar. The difference between 1 Gyr lookback time and the time of observation in Log sSFR shows more variation than in \sigsfr in all cases. 

UVCANDELS ID 05386 (Figure~\ref{fig:rad05386}) is a galaxy pair. The “center’’ is set at the brighter galaxy, but the signal becomes confused with radius as both galaxies contribute. The central region of the primary galaxy shows trends consistent with many young stars that may have formed a Gyr or more before observation. However, while the $t_x$ fractions hint at this, the differences in \sigsfr and Log sSFR are too small for firm conclusions.

UVCANDELS ID 08218 (Figure~\ref{fig:rad08218}) shows little evolution in \sigsfr and a roughly uniform decline in Log sSFR from 1 Gyr lookback time to the time of observation. The galactic center appears older yet remains consistently star forming.

Figure~\ref{fig:rad10335} shows UVCANDELS ID 10335, with a bright twisted outer arm and a faint upper feature. In the projected radial profiles this feature appears as a clump of low \sigsfr but relatively high Log sSFR, with both \sigsfr and Log sSFR declining fairly uniformly over the past 1 Gyr.

While some galaxies show a radius-independent decline in star formation, 08062 (Figure~\ref{fig:rad08062}) exhibits changes in Log sSFR suggesting that their \textit{outskirts} may be beginning to quench.

UVCANDELS ID 20524 is shown in Figure~\ref{fig:rad20524}. The apparent crossing within the inner 5 kpc may be an artifact of projection from the edge-on orientation. The overall profile is intriguing, but the edge-on view and lack of kinematic data prevent strong conclusions from photometry alone.
Figure~\ref{fig:rad22456} shows UVCANDELS ID 22456, an inclined spiral galaxy with arm knots visible in the F435W and F275W bands. The projected radial profiles of \sigsfr and log sSFR change little between 1 Gyr lookback time and the time of observation, aside from an overall decline in star formation. The log sSFR profile suggests this decline is nearly uniform, but somewhat stronger near the center than in the outskirts, leaving the outer regions mildly more star-forming. Although not statistically significant, this may hint at the earliest stages of inside-out quenching.

Figure~\ref{fig:rad23475} shows the radial profiles of UVCANDELS ID 23575. The \sigsfr, $A_v$, and $t_x$ fractions have similar radial shapes. \sigsfr shows little evolution between a 1 Gyr lookback time and the time of observation, but the projected Log sSFR profile declines overall with time.
\vspace{5cm}



\vfill

\section{UVCANDELS ID 02042}
UVCANDELS ID 02042 was initially selected by eye, but has not been included in the sample for analysis due to its high redshift making it an outlier compared to the rest of our sample that lies at a redshift of $<0.3$. While we have excluded it with particular concern as to how it may bias analysis that examines comparison with regions from the whole sample such as in Section~\ref{sec:global}, though its inclusion in this section did not significantly change our results. This galaxy initially was selected by eye and redshift was not a selection criterion, and while it is still a good candidate for resolved SED fitting, the drastically different redshift as compared to the sample necessitates separate analysis. 

\begin{figure}[H]
    \centering
    \includegraphics[width=0.9\linewidth]{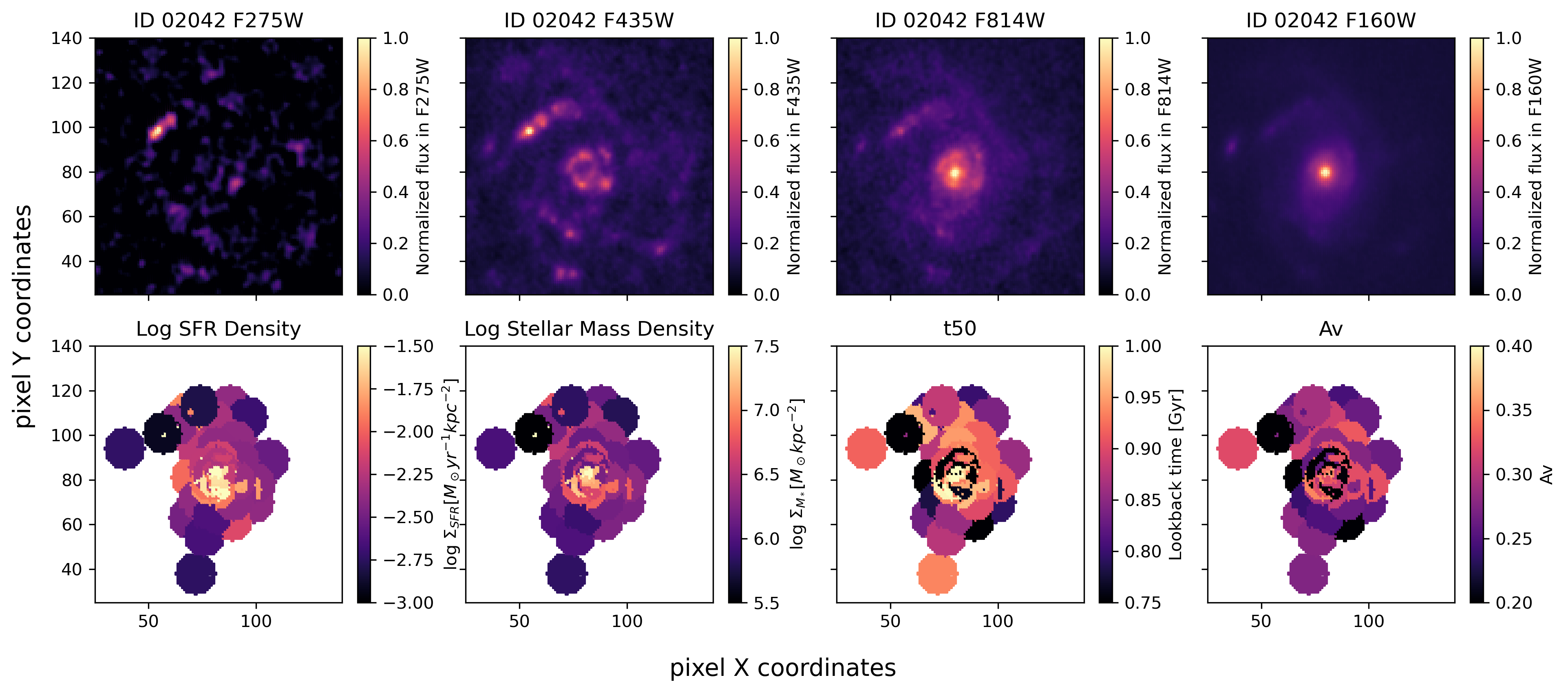}
    \caption{Postage stamps and property maps of UVCANDELS ID 02042 in pixel coordinates are shown. The segmentation map captures many key features seen in the F435W image, but the bright arm in the upper left appears only as small regions in the segmentation. These regions nonetheless show elevated SFR and stellar mass surface density relative to their surroundings, and the $t_{50}$ maps indicate they formed their stars at different times than nearby regions.}
    \label{fig:multi_02042}
\end{figure}

Though not included in our sample, 02042 is an interesting standalone example and we believe the results of the analysis performed on this galaxy merit inclusion. 
Figure~\ref{fig:multi_02042} shows postage stamps of UVCANDELS 02042. The galaxy combines extended low-SNR regions with compact areas that meet the SNR cutoff, most clearly in the bright structures in the upper-left quadrant. Although small in the segmentation map, these regions have elevated $\Sigma_{SFR}$ and $\Sigma_{M_*}$ compared to their surroundings, hosting some of the highest SFR and stellar mass densities in the galaxy. The $t_{50}$ segmentation map further indicates that these regions assembled their stellar mass at different times.

\begin{figure}
    \centering
    \includegraphics[width=0.9\linewidth]{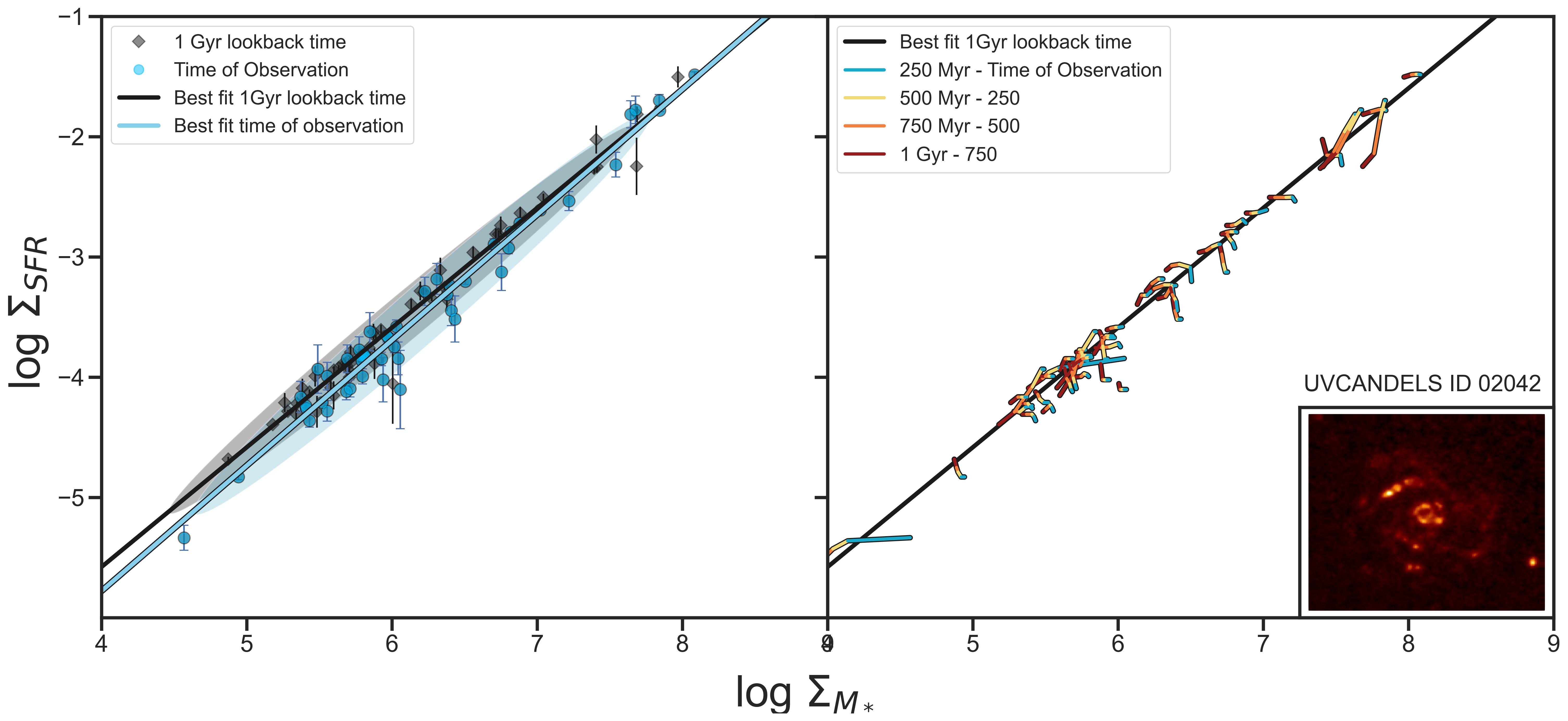}
    \caption{Similar to Figure~\ref{fig:sfr_mstar_lownum} the evolving \sigsfr-\sigmstar correlation of UVCANDELS 02042 has fewer regions and shows a sparser relation. The variety in evolutionary paths is clear in the right panel.}
    \label{fig:sfr_mstar_02042}
\end{figure}
\begin{figure}
    \centering
    \includegraphics[width=0.6\linewidth]{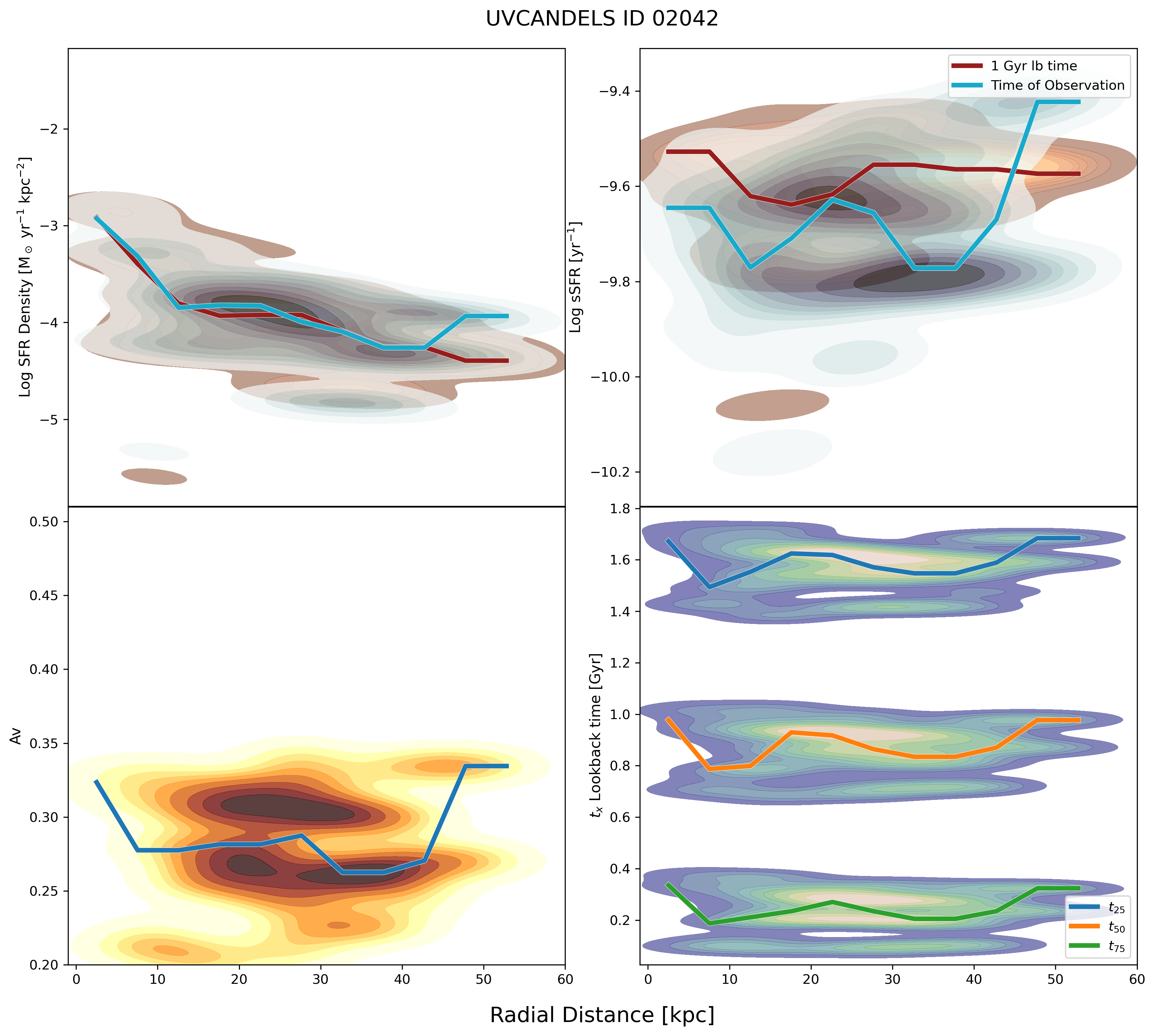}
    \caption{The projected radial profiles of UVCANDELS ID 02042 resemble those of other galaxies near the galactic center, with a peak in \sigsfr at both epochs that declines steeply with radius. In the outskirts, both \sigmstar and \sigsfr at the time of observation exceed their values at 1 Gyr lookback time, especially in \sigsfr around 50 kpc. This bright arc of star formation is visible in the images, and the SFHs indicate that this burst is likely recent}
    \label{fig:rad02042}
\end{figure}

In Figure~\ref{fig:sfr_mstar_02042} we see a slight increase in slope and decrease in normalization in the \sigsfr-\sigmstar correlation over the past 1 Gyr, but individual regions show diverse evolutionary paths. The radial profiles of UVCANDELS 02042 (Figure~\ref{fig:rad02042}) show a slight increase in \sigsfr and Log sSFR in the outskirts relative to 1 Gyr ago, suggesting that the bright outer-arm features are very recent. Dust peaks at both the center and outskirts, and radial growth appears fairly uniform.



\end{document}